\documentclass[preprint1,authoryear]{aastex}
\usepackage[]{graphicx}
\usepackage{epsfig}
\usepackage{hyperref}
\usepackage{natbib}
\newcommand{\beq}{\begin{equation}}
\newcommand{\eeq}{\end{equation}}
\newcommand{\nbeq}{\begin{equation*}}
\newcommand{\neeq}{\end{equation*}}

\shorttitle{Imaging SN1987A at 36 GHz}
\shortauthors{Potter et al.}

\begin{document}


\title{High resolution 36 GHz imaging of the Supernova Remnant of SN1987A}


\author{T. M. Potter\altaffilmark{1}, L. Staveley-Smith\altaffilmark{1,a},  C.-Y. Ng\altaffilmark{2}, Lewis Ball\altaffilmark{3}, B. M. Gaensler\altaffilmark{2,b}, M. J. Kesteven\altaffilmark{3}, R. N. Manchester\altaffilmark{3}, A. K. Tzioumis\altaffilmark{3} and G. Zanardo\altaffilmark{1}.}
\email{T.Potter@aip.org.au}


\altaffiltext{1}{International Centre for Radio Astronomy Research (ICRAR), School of Physics M013, The University of Western Australia, Crawley WA 6009, Australia.}
\altaffiltext{2}{Sydney Institute for Astronomy (SIfA), School of Physics, The University of Sydney, Sydney NSW 2006, Australia.}
\altaffiltext{3}{Australia Telescope National Facility (ATNF), CSIRO, PO Box 76, Epping NSW 1710, Australia.}

\altaffiltext{a}{Premier's Fellow in Radio Astronomy.}
\altaffiltext{b}{Federation Fellow, Australian Research Council.}
\altaffiltext{c}{CSIRO Fellow.}
%\altaffiltext{3}{present address: Center for Astrophysics}


\begin{abstract}
The aftermath of supernova (SN) 1987A continues to provide spectacular
insights into the interaction between a SN blastwave and its circumstellar
environment. We here present 36~GHz observations from the Australia
Telescope Compact Array of the radio remnant of SN~1987A. These new
images, taken in 2008 Apr and 2008 Oct, substantially extend the frequency
range of an ongoing monitoring and imaging program conducted
between 1.4 and 20~GHz. Our 36.2~GHz images have a diffraction-limited angular resolution of
$0\hbox{$.\!\!^{\prime\prime}$}3 - 0\hbox{$.\!\!^{\prime\prime}$}4$, which covers the gap between high resolution, low dynamic range VLBI images of the remnant and low resolution, high dynamic range images at frequencies between 1 and 20~GHz. The radio morphology of the remnant at 36~GHz is an elliptical ring with enhanced emission on the eastern and western sides, similar to that seen previously at lower frequencies. Model fits to the data in the Fourier domain show that the emitting region is consistent with a thick inclined torus of mean radius $0\farcs85$, and a 2008 Oct  flux density of $27\pm6$~mJy at 36.2~GHz. The spectral index for the remnant at this epoch, determined between 1.4~GHz and 36.2~GHz, is $\alpha=-0.83$. There is tentative evidence for an unresolved central source with flatter spectral index.

\end{abstract}





\section{Introduction}

The type IIP supernova 1987A is the most well-observed and published supernova in existence. As the brightest supernova seen since the invention of the telescope, it has been observed right across the electromagnetic spectrum from radio through to gamma rays. It was also the first supernova for which we have observations of the progenitor, as well as the first and only supernova thus far to have been associated with a neutrino burst event \citep{Hirata:1987p4127,Bionta:1987p13260}. Neutrino observations from SN1987A have been used to put a firm foundation on theories of core collapse and neutron star formation \citep{Janka:1997p13373} as well as giving a time of core collapse at 1987 February 23, 07:35 UT. Shortly after core collapse an expanding shockwave tore through the progenitor's stellar envelope. The resulting shock breakout resulted in a UV flash that ionized material in the circumstellar environment \citep{Luo:1991p372}, illuminating a central equatorial ring and two outer rings that we now know are part of a much larger hourglass structure \citep{Sugerman:2005p11362}. As yet, the mechanism for ring formation is not properly understood, however two competing theories, the interacting winds model \citep{Blondin:1993p14977,Martin:1995p11941,Tanaka:2002p19} and the binary merger model \citep{Podsiadlowski:2007p9791} offer plausible mechanisms for shaping the rings. At present the shock is transiting the equatorial ring and continues to provide insights into the hydrodynamics of a unique young supernova remnant as followed by X-ray, optical and radio telescopes. 

In the radio SN1987A was imaged as soon as a few days after detection. On 1987 Feb 25, \citet{Turtle:1987p2190} detected emission at 843~MHz with the Molonglo Observatory Synthesis Telescope (MOST) along with detections at 1.4, 2.3 and 8.4~GHz by an interferometer formed between the Parkes 64~m and Tidbinbilla 34~m antennas. Emission at 843~MHz peaked around 130~mJy on 1987 Feb 26-27. Just over five days after the neutrino burst (1987 Feb 28), \citet{Jauncey:1988p3319} used VLBI techniques to resolve the expanding radiosphere. Continued observations  saw the supernova fade from its intial radio burst to almost undetectable levels within a few months \citep{Turtle:1987p2190}. From this point on, radio emission from SN1987A deviated radically from the trend of typical type IIP supernova light curves. Radio emission from the remnant returned in 1990 with a re-detection at 843~MHz by MOST on July 5 \citep{Ball:2001p6811} and the six 22m antennas of the Australia Telescope Compact Array (ATCA) a month later \citep{StaveleySmith:1992p680}. This has been interpreted as a collision between the expanding shock front and an ionized hydrogen region about half the distance to the optical ring \citep{Chevalier:1995p4450}. Since then, an ongoing observing campaign \citep{Gaensler:1997p7998,Manchester:2002p1,Manchester:2005p7378,StaveleySmith:2007p2,Gaensler:2007p42} has followed radio emission from the developing remnant at frequencies ranging from 1.4 to 20 GHz. This saw the radio emission rise at monotonically increasing rate as the shockwave began its crossing of the ring \citep{zanardo:2009}. Such behaviour puts the remnant of SN1987A in the somewhat rare category of young and brightening supernova remnants, along with the Galactic supernova remnant G1$\cdot$9+0.3 \citep{Reynolds:2008p14576,Murphy:2008p14040}.  

High precision images of the remnant at high resolution are important for studying the morphology and evolution of the blastwave interaction. Previous estimates of the optical ring geometry \citep{Plait:1995p25} have shown that it is an ellipse with semi-major axis $0\farcs858\pm0\farcs011$ and semi-minor axis $0\farcs621 \pm 0\farcs011$, thus requiring very high resolution imaging. Cameras aboard the Hubble Space Telescope (HST) are capable of imaging at $0\farcs03 - 0\farcs2$ pixel$^{-1}$  \citep{biretta:2008p10404} which has enabled detailed imaging of the structure of the equatorial ring as well as the development of the reverse shock \citep{Michael:2003p28}. Pixels on the ACIS detector aboard the Chandra X-ray Observatory register half the total encircled energy of incident photons within a circle of diameter 0\farcs8 [half power beam diameter (HPD)], although this can be improved to around $0\farcs6 $ HPD with subsampling techniques \citep{Tsunemi:2001p8231}. In 2007 October, \citet{tingay:2009} used an e-VLBI array formed between the ATCA, Parkes, and Mopra (22m) antennas to image the remnant of SN1987A at 1.4~GHz with a restoring beam of 85x168 milli-arcseconds. However the limited range of baselines made such an array insensitive to details larger than $0\farcs4$. In the radio the ATCA is capable of resolving details as small as 0\farcs2 at around 40 GHz, which nicely fills in the gap left by the VLBI observation and images at the lower frequencies. Imaging at higher frequencies around 100~GHz is also possible, although this capability is enable to for only five of the six antennas, cutting the number of baselines by a third as well as reducing the maximum baseline length to 3~km. This translates to roughly the same resolution (approximately $0\farcs2$) as the 40~GHz band.  Unfortunately at higher frequencies atmospheric opacity degrades the image with approximately an order of magnitude higher effective system temperature \citep{atnf-7mm}. With an aim to obtain the highest resolution images currently possible with the ATCA, the ongoing imaging observations of \citet{YNg:2008p14427} were extended to include frequencies around 36~GHz. In this paper we present the latest 36~GHz data from the recent extension to these ongoing observations.

\section{Observations \& reduction}\label{obsreduc}

Our observations of the supernova remnant were made with the ATCA\footnote[1]{The Australia Telescope Compact Array is part of the Australia Telescope which is funded by the Commonwealth of Australia for operation as a National Facility managed by CSIRO.} at center frequencies of 34.88 and 37.44 GHz on 2008 April 25 and 2008 October 7,8,12 using antenna configuration 6A with multifrequency imaging over $2\times128$ MHz bandwidth in all Stokes parameters. Atmospheric phase stability at the ATCA was monitored with two 1.8 m antennas spaced 230 m apart designed to measure the path noise of a 30~GHz signal from the geostationary Optus-B3 communications satellite. A path noise of below 300 $\mu$m is considered good for millimetre-wave astronomy at the ATCA. Table \ref{obscond} shows observing conditions and un-flagged uv coverage for each observation. 

\begin{small}
\begin{table}[h]
\begin{center}
\caption{Observing conditions for 36.2~GHz observations of SN1987A. \newline}\label{obscond}
\begin{tabular}{lcc}
\\
\tableline \hline
 Date & Path noise & Integration time\\
& ($\mu$m) &  (hr)  \\
\tableline
 2008 Apr 25 & $334 \pm 133$ & $11.3$  \\
 2008 Oct 7 & $131 \pm 22$ & $7.4$  \\
 2008 Oct 8 & $88 \pm 10$ & $3.9$  \\
 2008 Oct 12 & $184 \pm 60$ & $9.8$  \\
\tableline
\end{tabular}
\medskip\\
\end{center}
\end{table}
\end{small}

The 2008 Apr 25 observation was hampered by poor weather, consequently around 3 hours of uv-data was discarded. The secondary flux calibrator PKS 0637-752 was observed for 2 minutes in hourly intervals and the phase calibrator PKS 0530-727 was observed for 1.5 minutes per 6 minutes integration time on source. A flux density of $3.93 \pm 0.06$ and $3.83 \pm 0.06$ Jy at 34.88 and 37.44 GHz was obtained for the 2008 October observations of PKS 0637-752  by bootstrapping an observation of the primary flux calibrator Uranus at similar elevation  ($\approx35^{o}$) from the night of 2008 October 8. The flux density of PKS 0637-752 for the 2008 April 25 observation was similarly determined to be  $3.71 \pm 0.07$ and $3.61 \pm 0.08$ Jy using a later 2008 May 25 observation of both the calibrator and Uranus (at the shortest baseline). The phase center was located about 5\farcs5 south of the remnant to avoid defects \citep{ekers:1999}. We used the software package \emph{Miriad} \citep{Sault:1995p14420} to perform image reduction and analysis. Continuum images were made with robust=0.5 weighting \citep{Briggs:1995p8023} as this represented the best compromise between resolution and signal-to-noise ratio. Natural weighting (robust=2.0) was used in both the flux and polarisation measurements in order to preserve as much signal as possible. The \emph{CLEAN} algorithm \citep{Hogbom:1974p9252,Clark:1980p9245} was employed to deconvolve the resulting continuum images. Table \ref{restorbeam} gives the size and position angle of the restoring beam, as well as the rms noise of the resulting image in each case.
 
 \begin{table*}[h]
\begin{center}
\caption{Image parameters for the 2008 observations at 36.2~GHz. \newline}\label{restorbeam}
\begin{tabular}{lcccc}
\tableline \hline
 Date & Robust & Restoring beam & Position angle & Rms noise   \\
& & (\arcsec) &  (\arcdeg) & (mJy~beam$^{-1}$) \\
\tableline
 2008 Apr 25 & 0.5 & $0.34\times0.24$ & 45.0 & 0.10 \\
 & 2.0 & $0.38\times0.28$ & 44.6 & 0.10 \\
 2008 Oct 7-12 & 0.5 & $0.37\times0.22$ & -7.7 & 0.06 \\
& 2.0 & $0.43\times0.25$ & -8.3 & 0.06 \\
2008 Apr-Oct & 0.5 & $0.33\times0.24$ & -1.3 & 0.06 \\
& 2.0 & $0.38\times0.27$ & -1.0 & 0.06 \\
\tableline
\end{tabular}
\medskip\\
\end{center}
\end{table*}

\begin{figure}[h] 
\begin{center}
\includegraphics[width=6.5cm, angle=270]{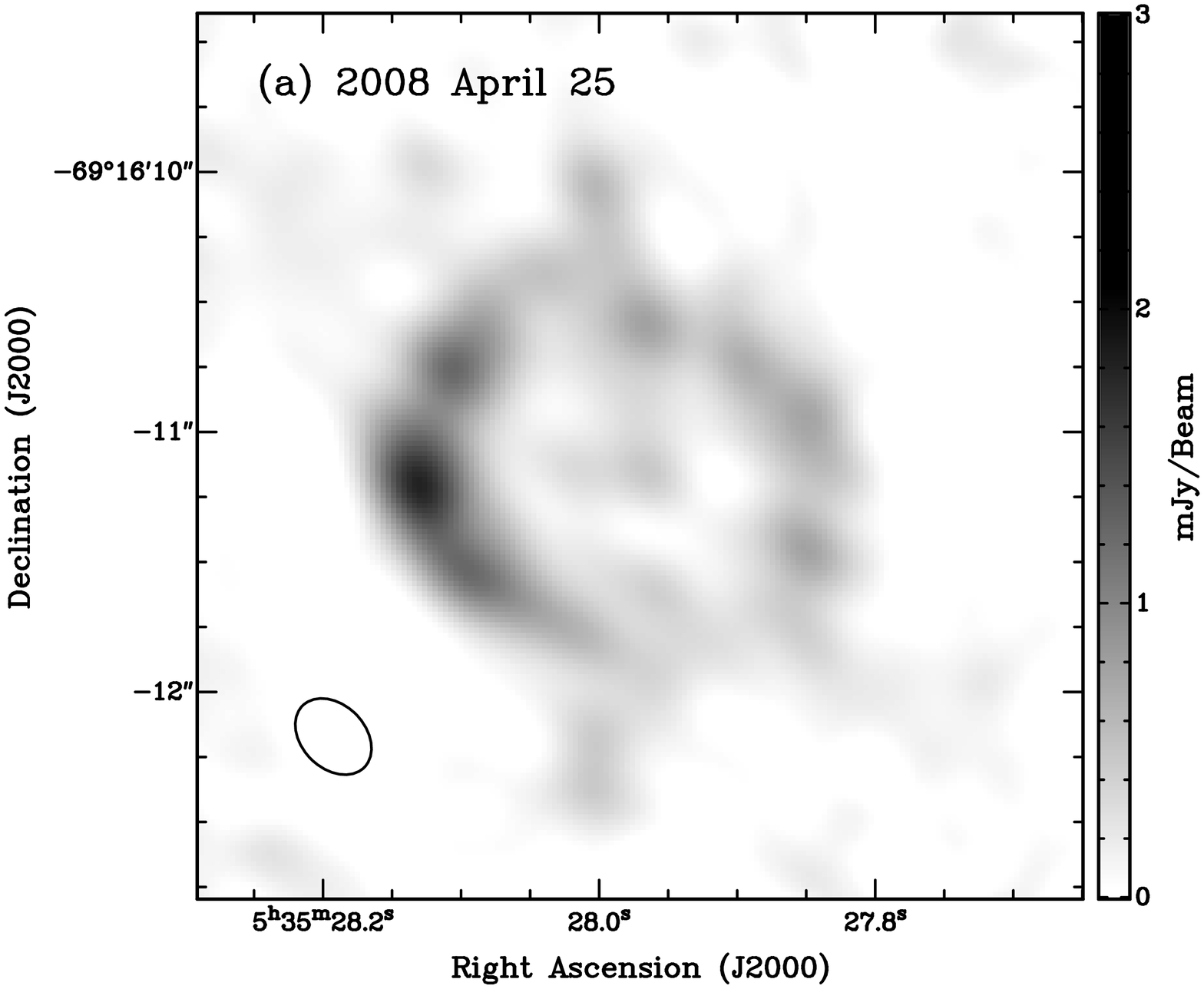} 
\includegraphics[width=6.5cm, angle=270]{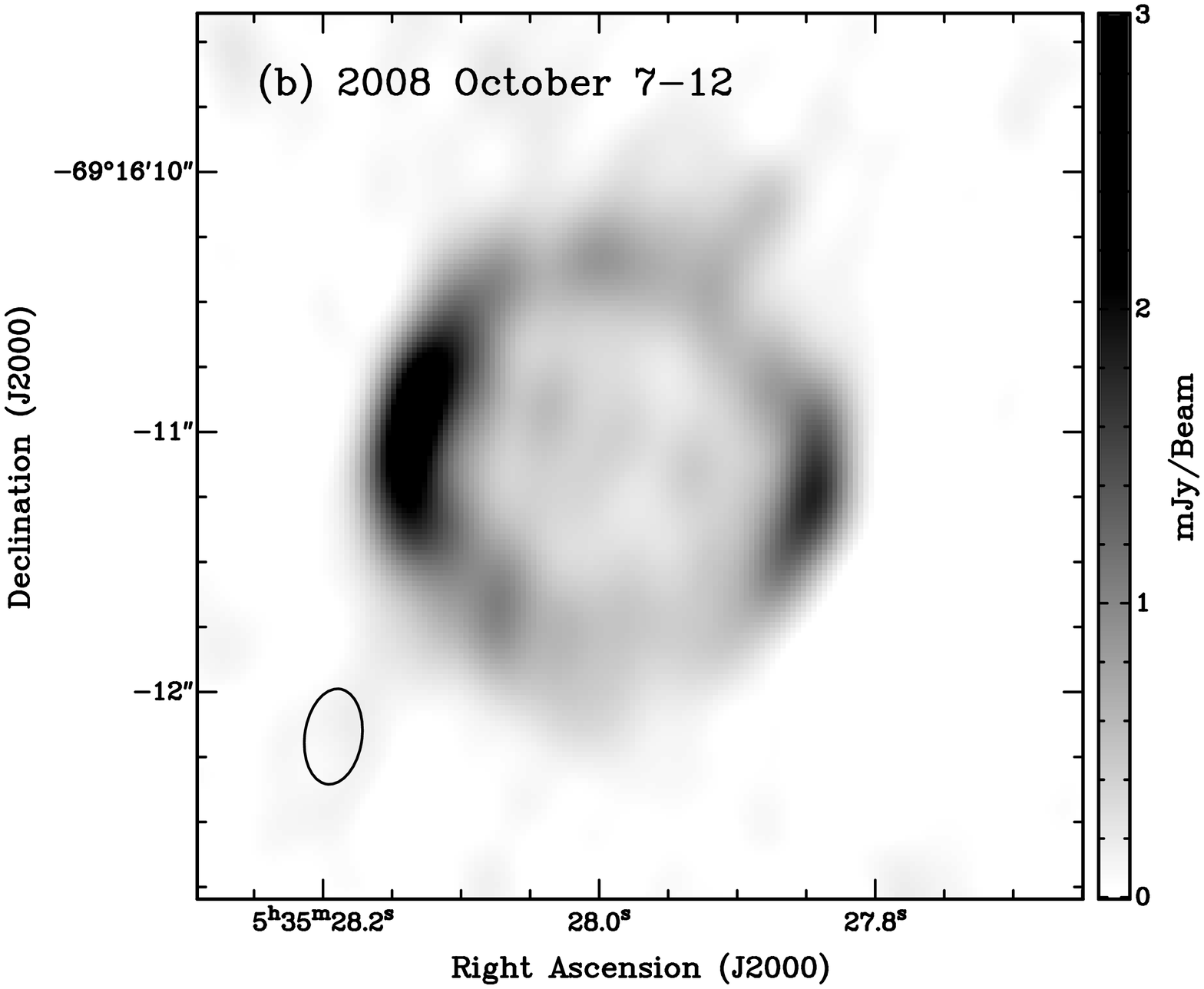}
\includegraphics[width=6.5cm, angle=270]{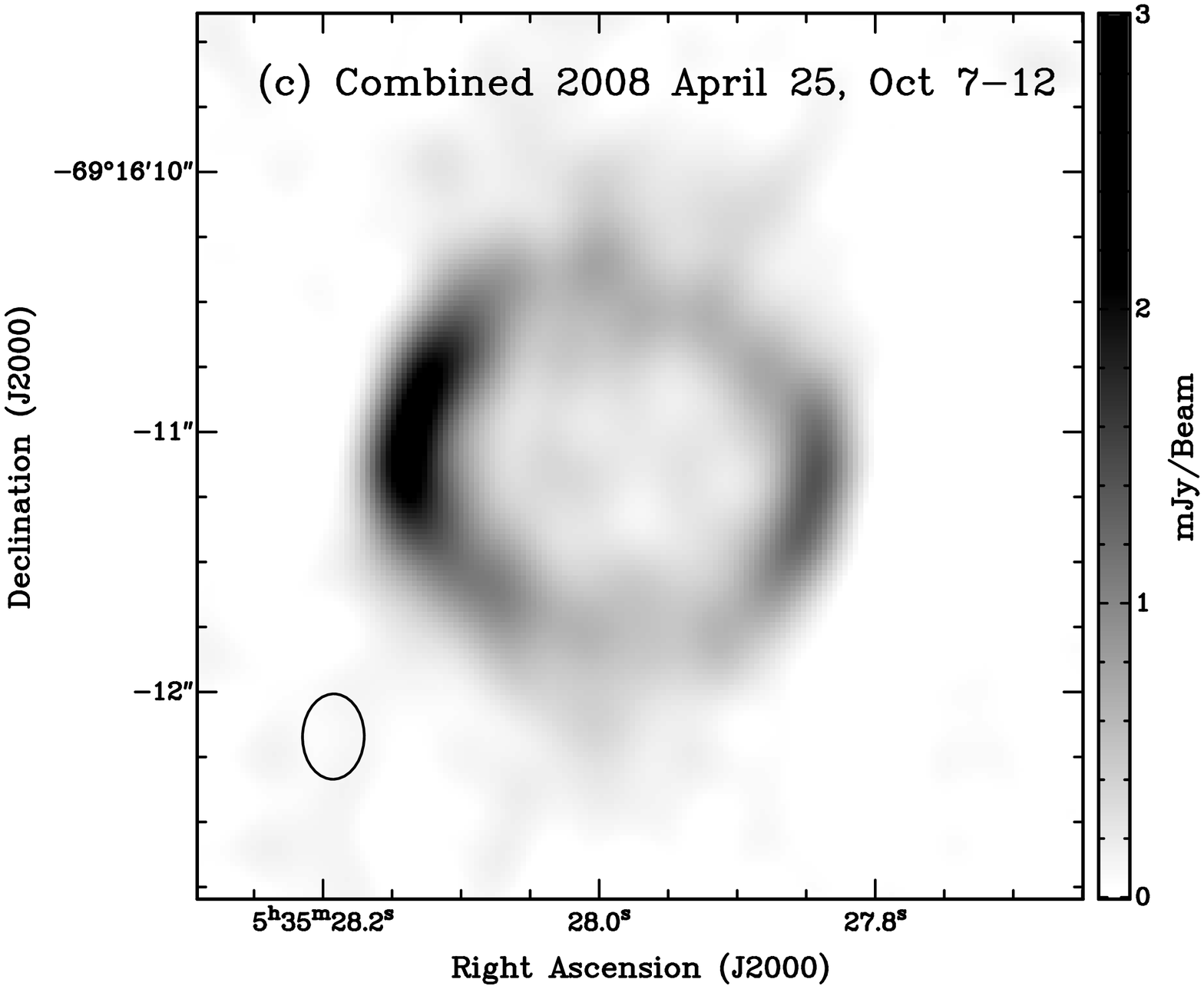}
\caption{Continuum images of SN1987A made with the ATCA at an average frequency of 36.2~GHz and robust=0.5 weighting: (a) 2008 April 25; (b) 2008 combined October data; and (c) 2008 combined observations. The 2008 April, October and combined images have an rms noise of 0.10, 0.06 and 0.06~mJy~beam$^{-1}$ respectively. The linear greyscale on each image ranges from 0 to 3~mJy~beam$^{-1}$. The black ellipse at lower left is the beam as given in Table \ref{restorbeam}.}\label{images}
\end{center}
\end{figure}

Shown in Figure \ref{images} are the deconvolved Stokes-I continuum images of the 36.2~GHz observations.  Figure \ref{images}(a) is the image from the 2008 April 25 observation, Figure \ref{images}(b) is the combined 2008 October image, and Figure \ref{images}(c) is the combined 2008 April and 2008 October images. Note that the approximately north-south extensions prominent in the images are probably artifacts caused by insufficient uv-coverage and increased phase errors when observing at low elevation. 

\clearpage

\clearpage

For purposes of comparison, we also obtained a contemporaneous observation at 8.6~GHz on 2008 Oct 9-10 as well as our regular monitoring at 1.4-8.6~GHz. The 8.6~GHz observation was reduced as described in \citet{YNg:2008p14427}, and the 1.4-8.6~GHz flux monitoring observations and reduction are as described in \citep{zanardo:2009} [see also \citet{Manchester:2002p1}]. 

\section{Results and discussion}

\subsection{Flux density and spectral index of the remnant.}

\begin{figure}[h] 
\begin{center}
\includegraphics[width=7.65cm, angle=0]{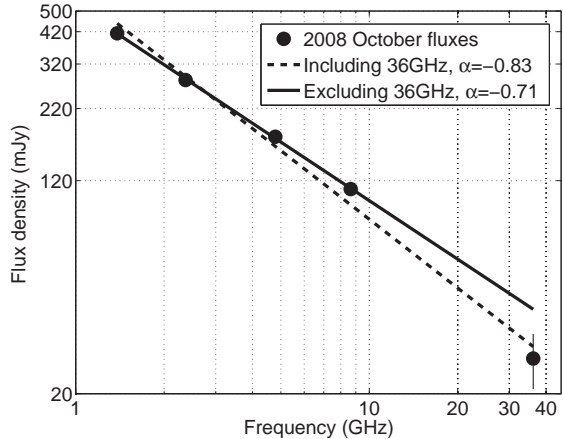} 
\caption{Spectral index of SN1987A as determined by a power-law fit to the flux densities of the 2008 October monitoring observations both with and without the 36.2~GHz observation. The spectral indices from the fit are $\alpha = -0.71$ for the solid line and $\alpha=-0.83$ for the dashed line.}\label{slopedetermine}
\end{center}
\end{figure}

A flux density estimate was obtained for the naturally weighted (robust=2.0) 2008 April and October images. The amplitudes and phases of the April observation were affected by prevailing weather conditions, making the 2008 April 25 flux density estimate unreliable. A flux density of $27 \pm 6$~mJy was measured for the combined October image by imaging using natural weighting and taking a cut $6\sigma$ above an rms noise of 60~$\mu$Jy beam$^{-1}$. 
The flux density of the 1.4, 2.4, 4.8, and 8.6~GHz 2008 October monitoring observations was measured to be $414 \pm 12, 280 \pm 8,173 \pm 6$, and $112 \pm 6$~mJy, respectively. This yields a spectral index $\alpha=-0.71 \pm 0.01$. When used to predict the 2008 October flux density  at 36.2~GHz, the fitted power spectrum predicts a flux density of $41 \pm 6$~mJy. The 2008 October flux density of $27 \pm 6$~mJy therefore lies 1.3$\sigma$ lower than the power law fit to the other frequencies. If we include the 2008 Oct observation in the fit, we obtain a spectral index of $\alpha=-0.83$, where $S_{\nu}\propto \nu^{\alpha}$. A comparison of the fits is shown in Figure \ref{slopedetermine}.

\subsection{Modeling in the Fourier domain and a limit on a central source.}\label{fourier_models}
 
\begin{figure*}[h]
\begin{center}$
\begin{array}{cc}
\includegraphics[width=4cm, angle=270]{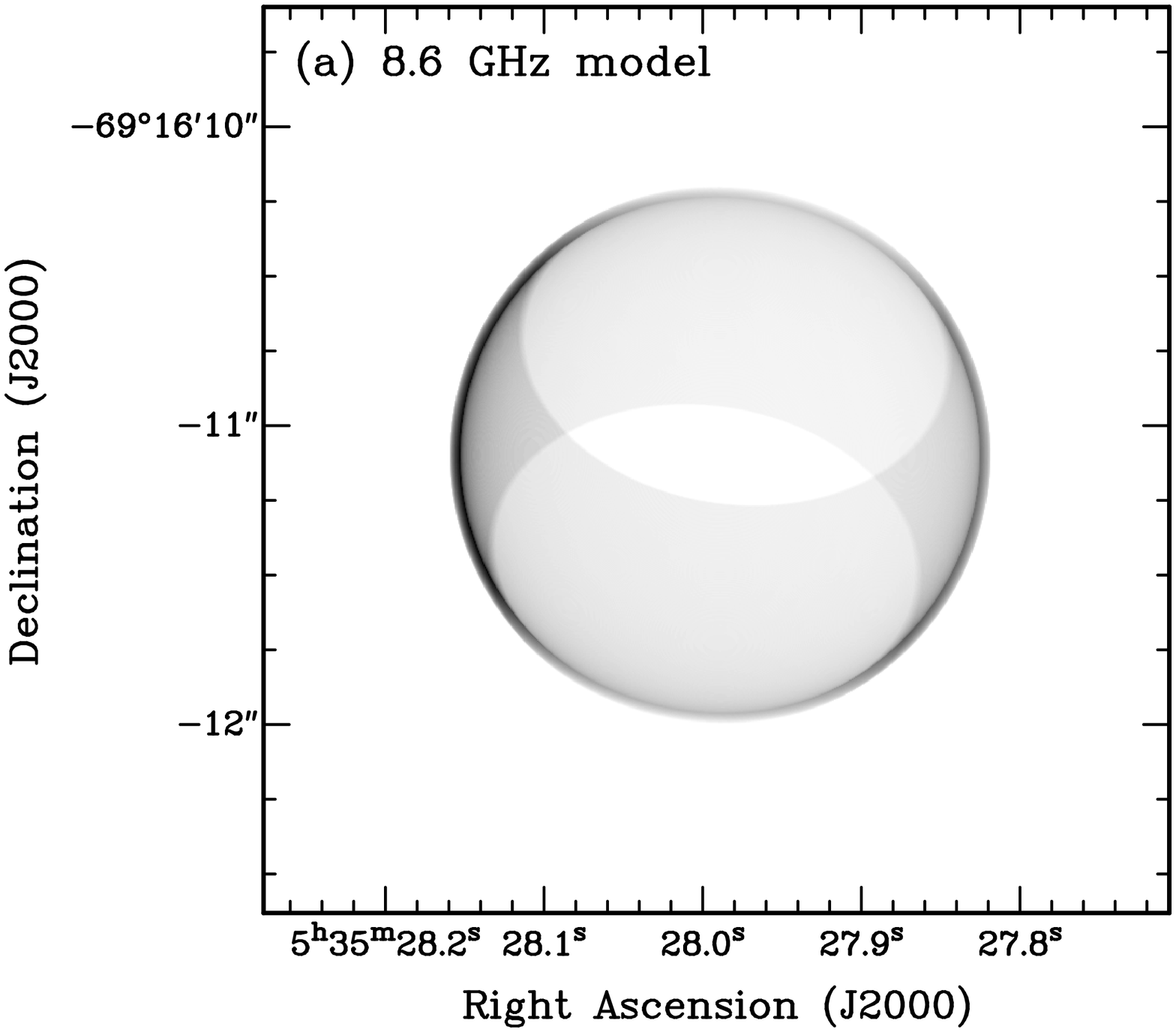} &
\includegraphics[width=4cm, angle=270]{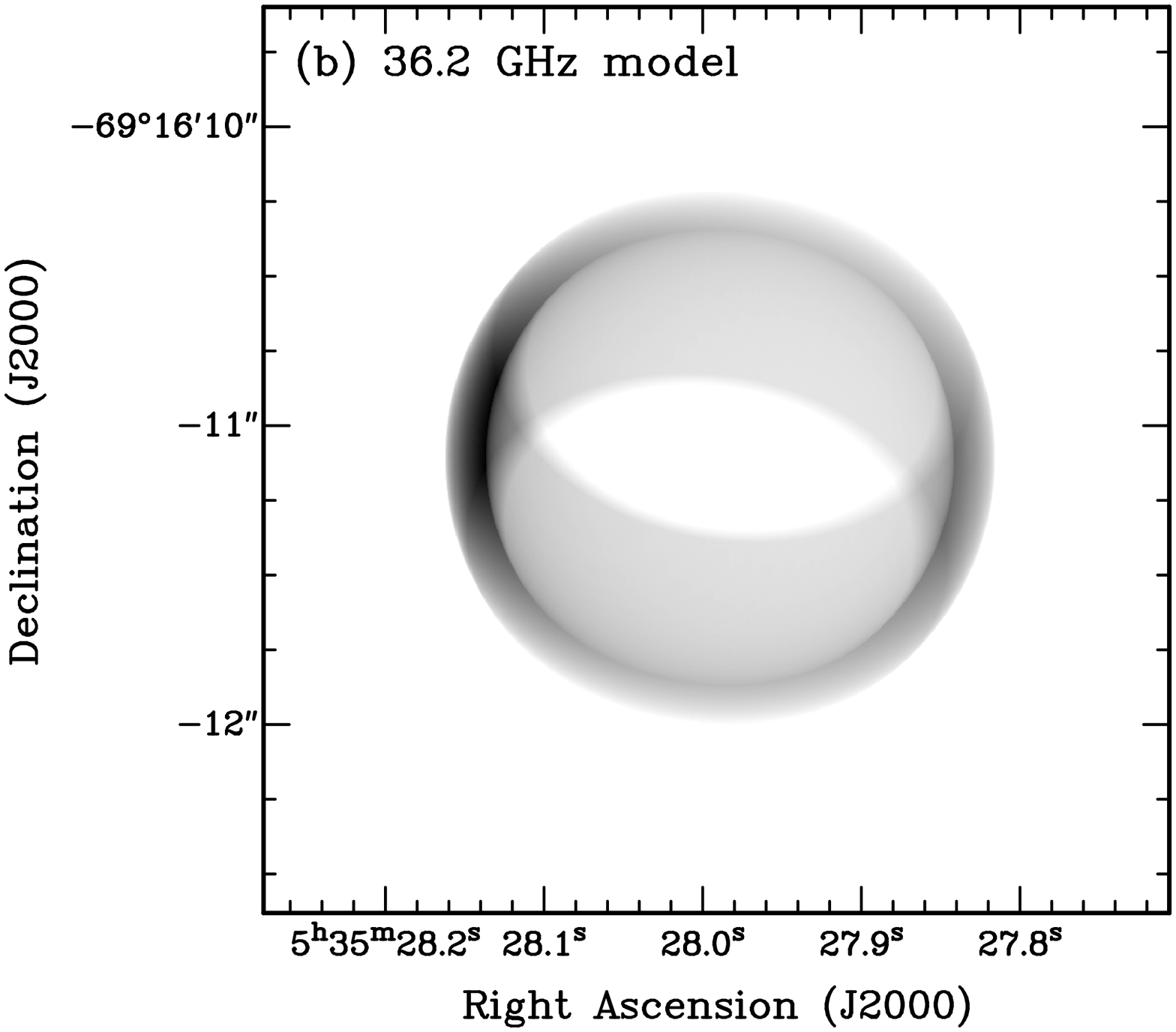} \\
\includegraphics[width=4cm, angle=270]{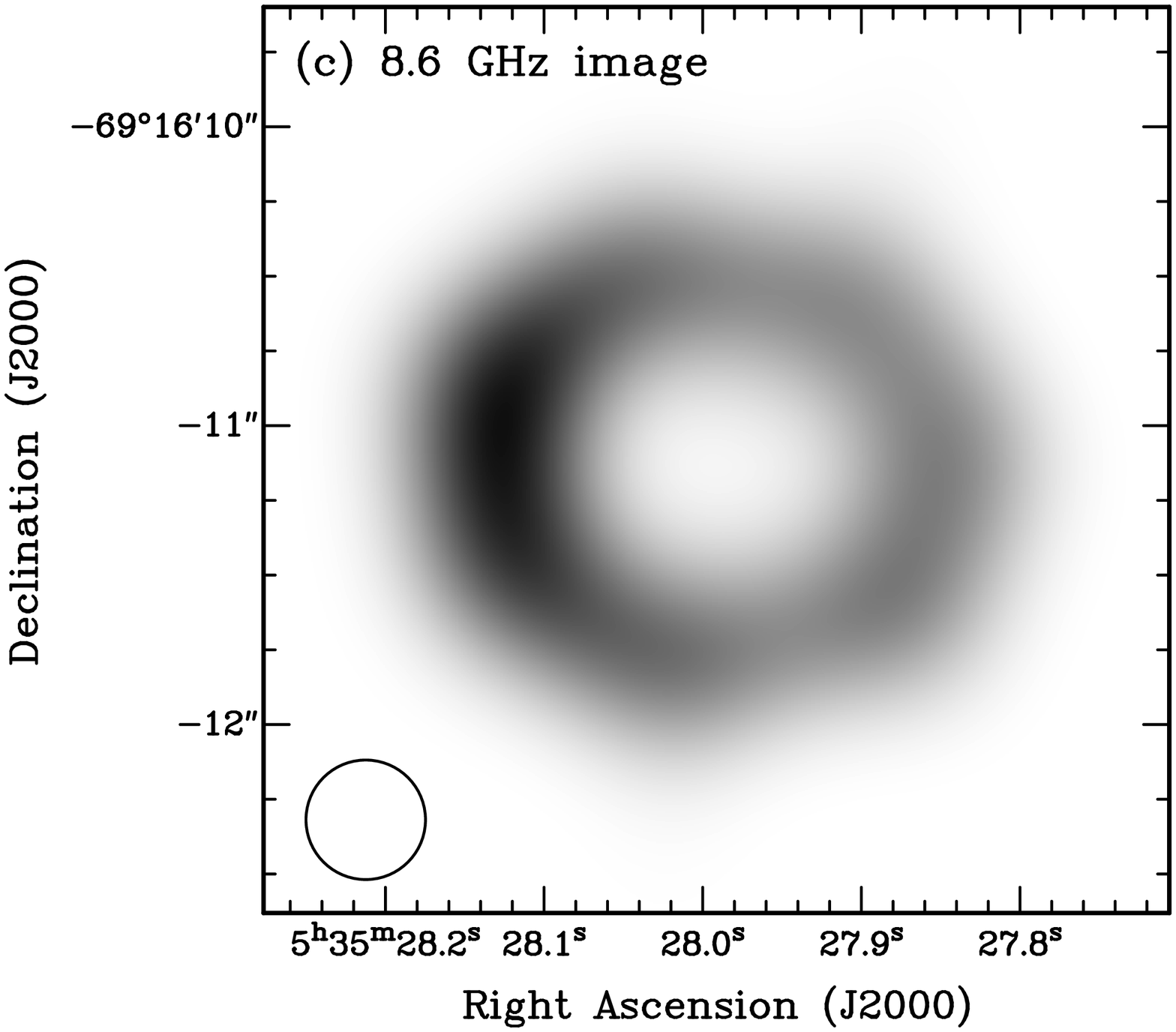} &
\includegraphics[width=4cm, angle=270]{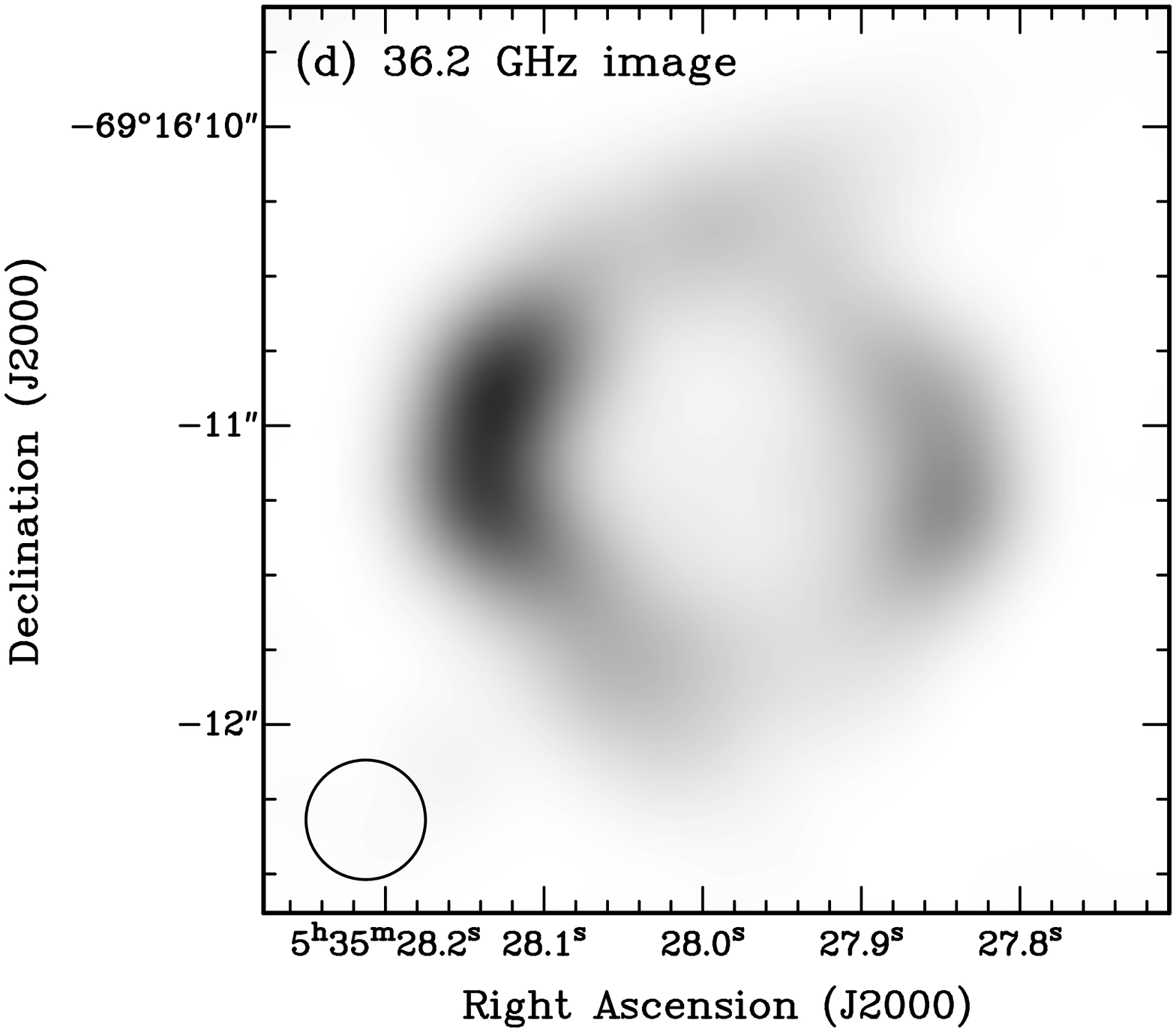} \\
\includegraphics[width=4cm, angle=270]{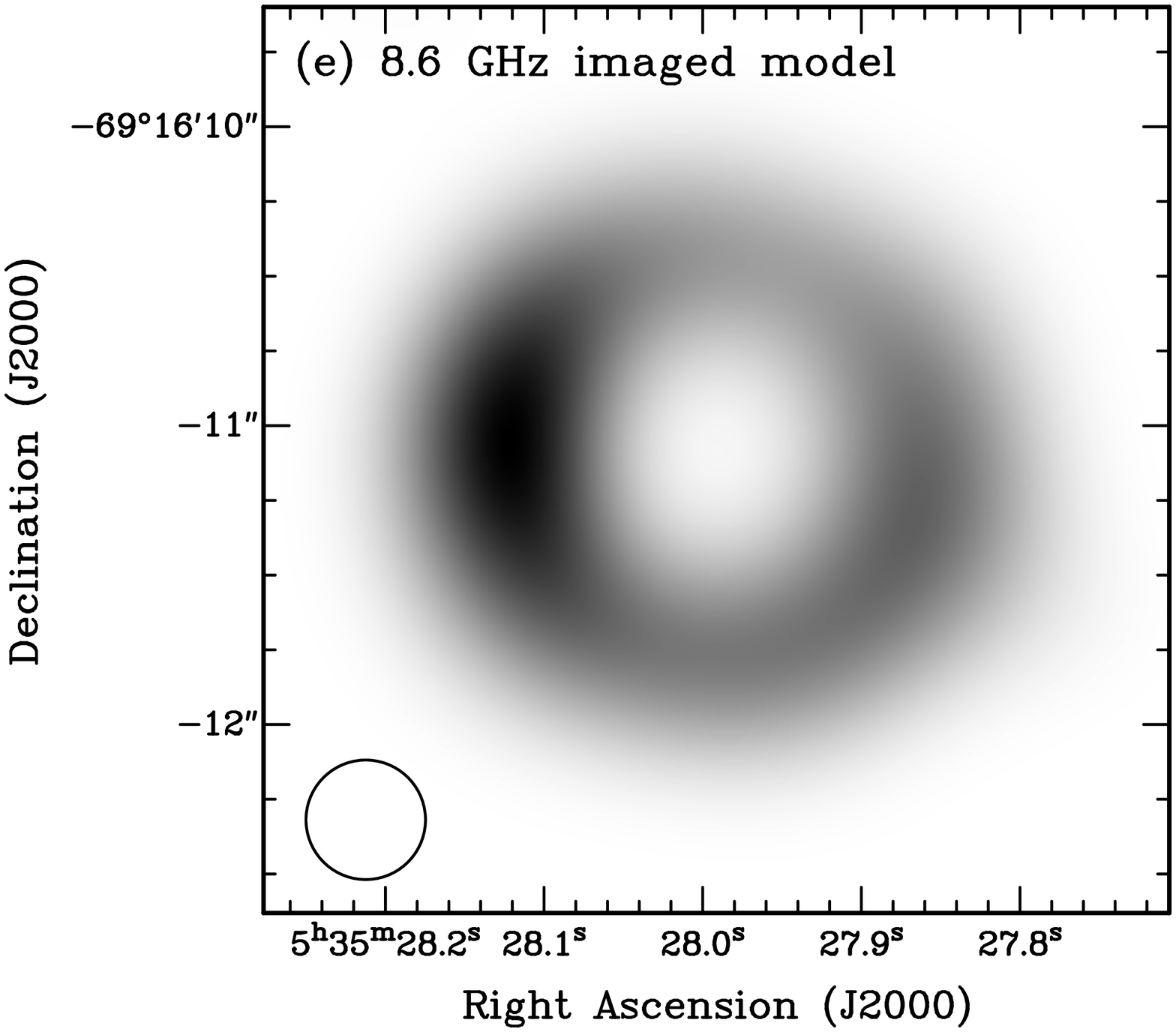} &
\includegraphics[width=4cm, angle=270]{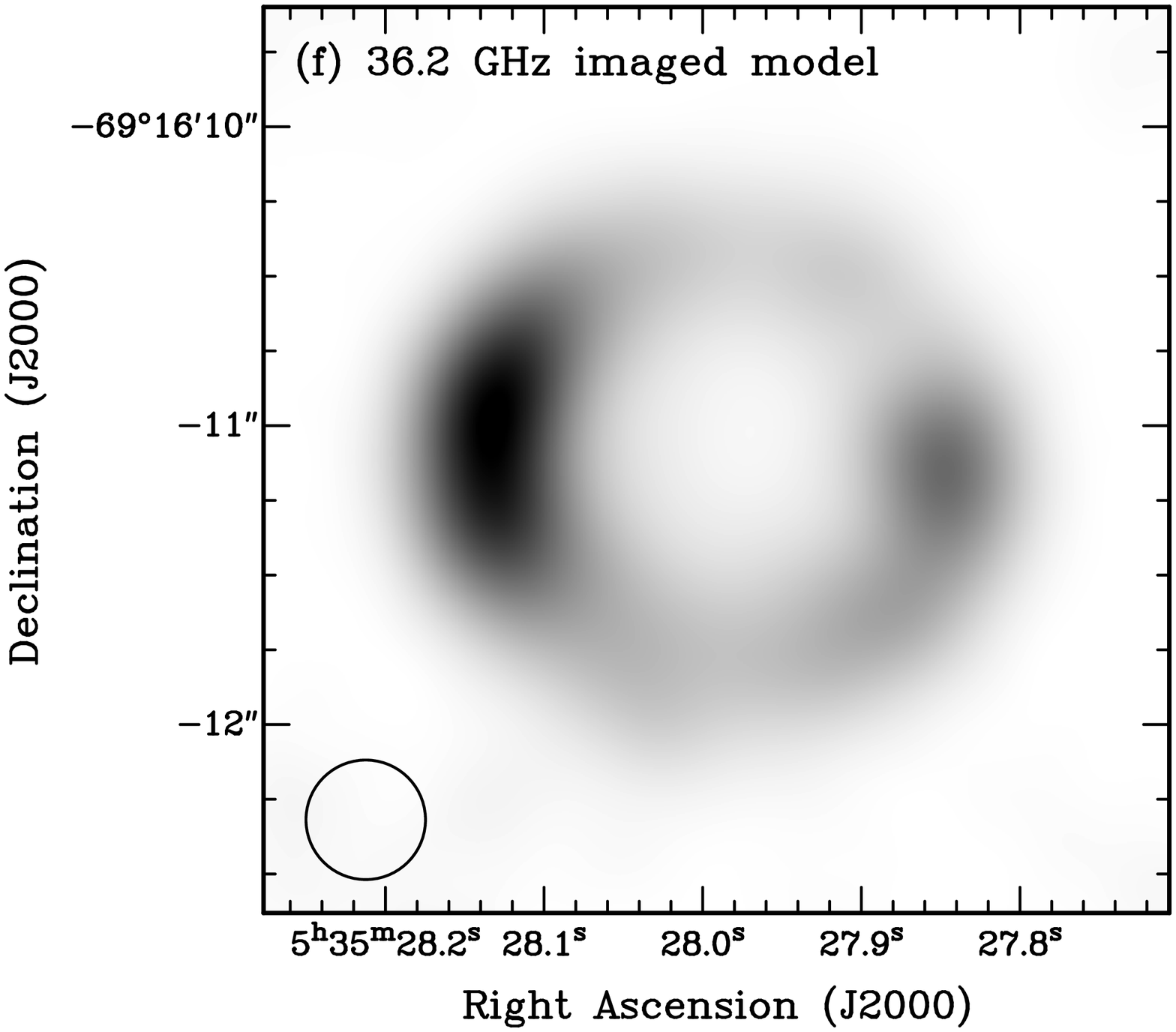} \\
\end{array}$
\end{center}
\caption{Fourier models and images from the 2008 October observation at 8.6~GHz and 36.2~GHz. The left column contains 8.6~GHz images, the right column 36.2~GHz images obtained using the same MAXEN deconvolution algorithm. In the top row are the truncated-shell models, the middle row contains the real images. The bottom row contains model images formed by inserting the model visibility data, generated from the top row, into the observed visibility data from the middle row and running them through the same imaging pipeline. The model images (bottom row) thus include noise and have the same uv-coverage as the real observations (middle row). All images have been restored using an identical $0\farcs4$ beam and have a linear greyscale that ranges from 0-13~mJy for the 8.6~GHz images and 0-3.5~mJy for the 36.2~GHz images respectively. }
\label{models_and_maps}
\end{figure*}

 We modeled the 2008 Oct 36.2 and 8.6~GHz data in the Fourier domain using a truncated shell model described by \citet{YNg:2008p14427}. The shell has a
radius $R$ and a half-opening angle $\theta$ (i.e., $\theta=90\arcdeg$
and 0\arcdeg\ corresponds to a spherical shell and an equatorial ring respectively), and it is inclined at 43.4\arcdeg\ south of the line of sight at a position angle -7.6\arcdeg (or 7.6\arcdeg clockwise) from north, as determined from optical observations \citep{pun:2007}. We modeled the east-west asymmetry by a linear gradient in the
equatorial plane, more details of the modeling can be refered to \citet{YNg:2008p14427}. The best fit model parameters at the two frequencies are shown in Table \ref{fourier_fit} and displayed graphically in the top row of Figure \ref{models_and_maps}. Both models are roughly consistent,  each with 30\% greater flux in the eastern lobe. The 8.6 and 36.2 GHz fitted shell widths do not agree at the $3\sigma$ level. If this is a physical effect, it may be due to a positional dependency of the distribution of particle energies in the forward shock. The small shell width characteristic of the fits to low frequency radio emission may correspond to a proportionately larger population of newly swept up particles at the forward shock. However shell thickness $\delta$ is a parameter that is not well constrained by the fit, as discussed in \citet{YNg:2008p14427} and demonstrated in Table 2 of that paper. 

 \begin{table*}[h]
\begin{center}
\caption{Best fit truncated-shell models of the 2008 Oct 36.2~GHz and 2008 October 8.6~GHz observations. \newline }\label{fourier_fit}
\begin{tabular}{lcccc}
\tableline \hline
Frequency  & Flux density & Mean shell radius & Shell width & Half-opening   \\
  &  $f$  & R & $\delta$   &  angle $\theta$     \\
  (GHz) &   (mJy) &  (\arcsec) &  (\%)  &  (\arcdeg)   \\
\tableline
36.2~GHz & 32.7 & $0.850 \pm$ 0.060 & $16 \pm 4$ & $29.2 \pm 2.0$  \\
8.6~GHz & 108.0 & $0.892 \pm 0.002$ & $0.04 \pm 2.6$ & $35.6 \pm 0.6$ \\
\tableline
\end{tabular}
\medskip\\
\end{center}
\end{table*}

\clearpage

\begin{figure}[h]
\begin{center}
\includegraphics[width=6.0cm, angle=270]{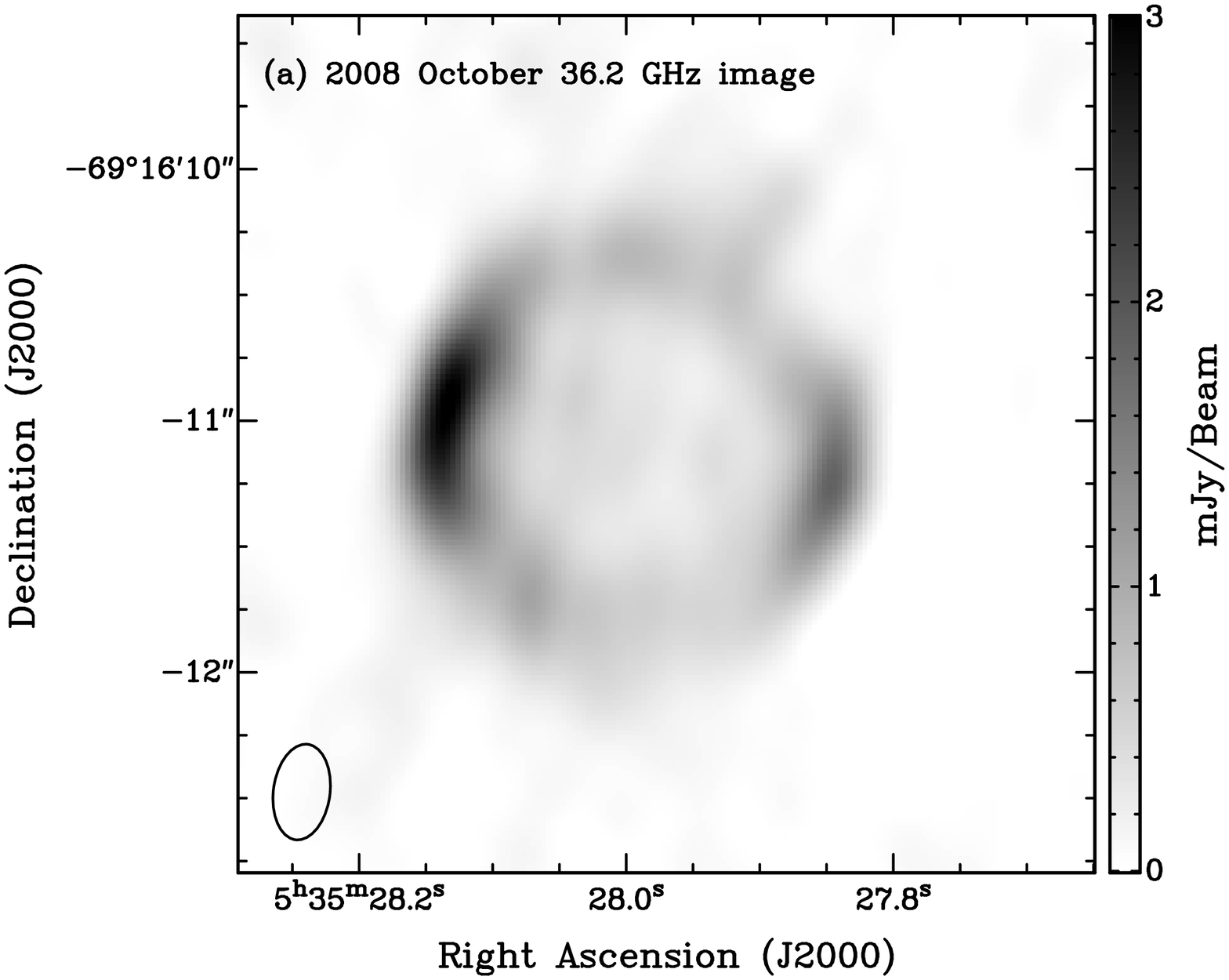}
\includegraphics[width=6.0cm, angle=270]{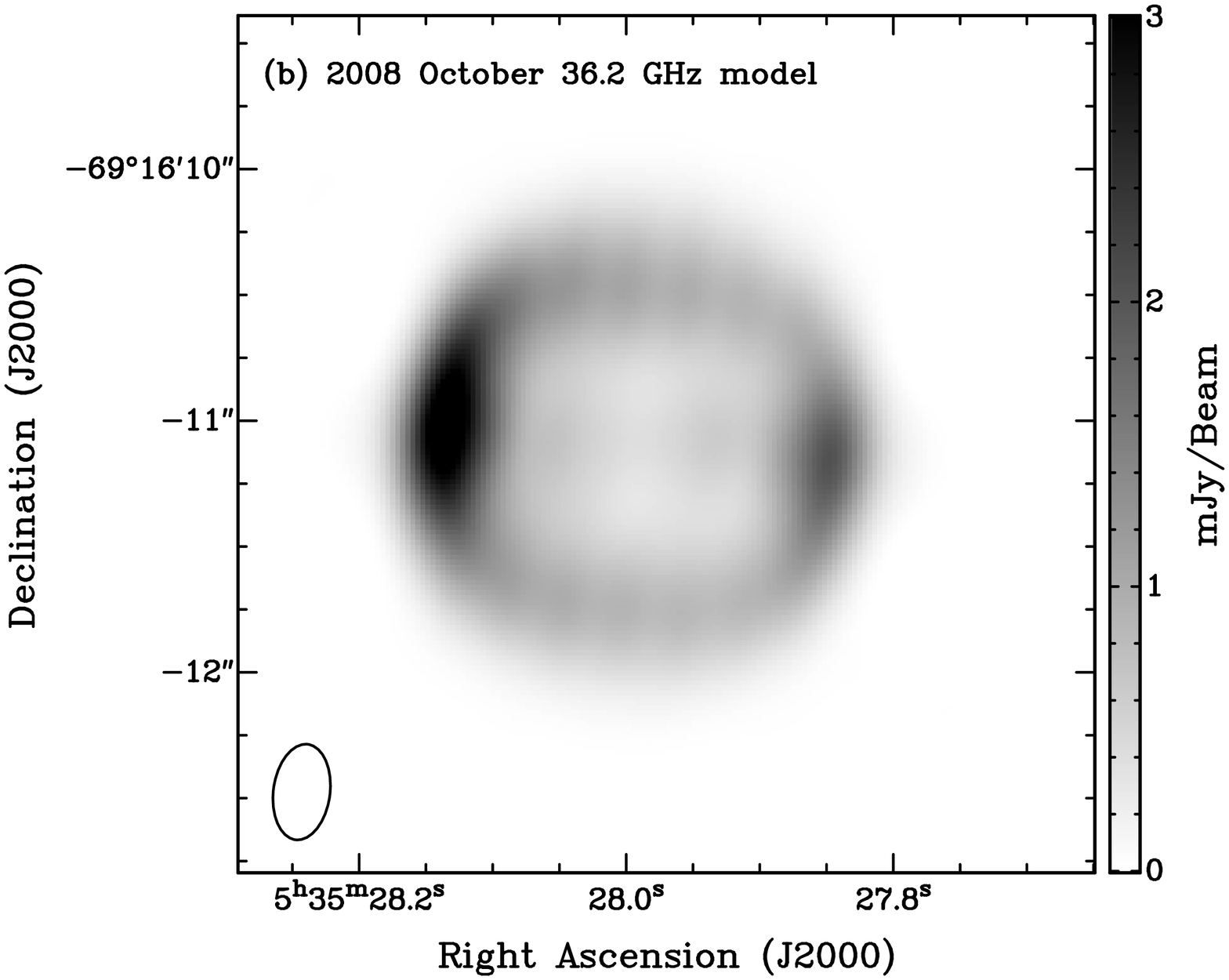}
\includegraphics[width=6.0cm, angle=270]{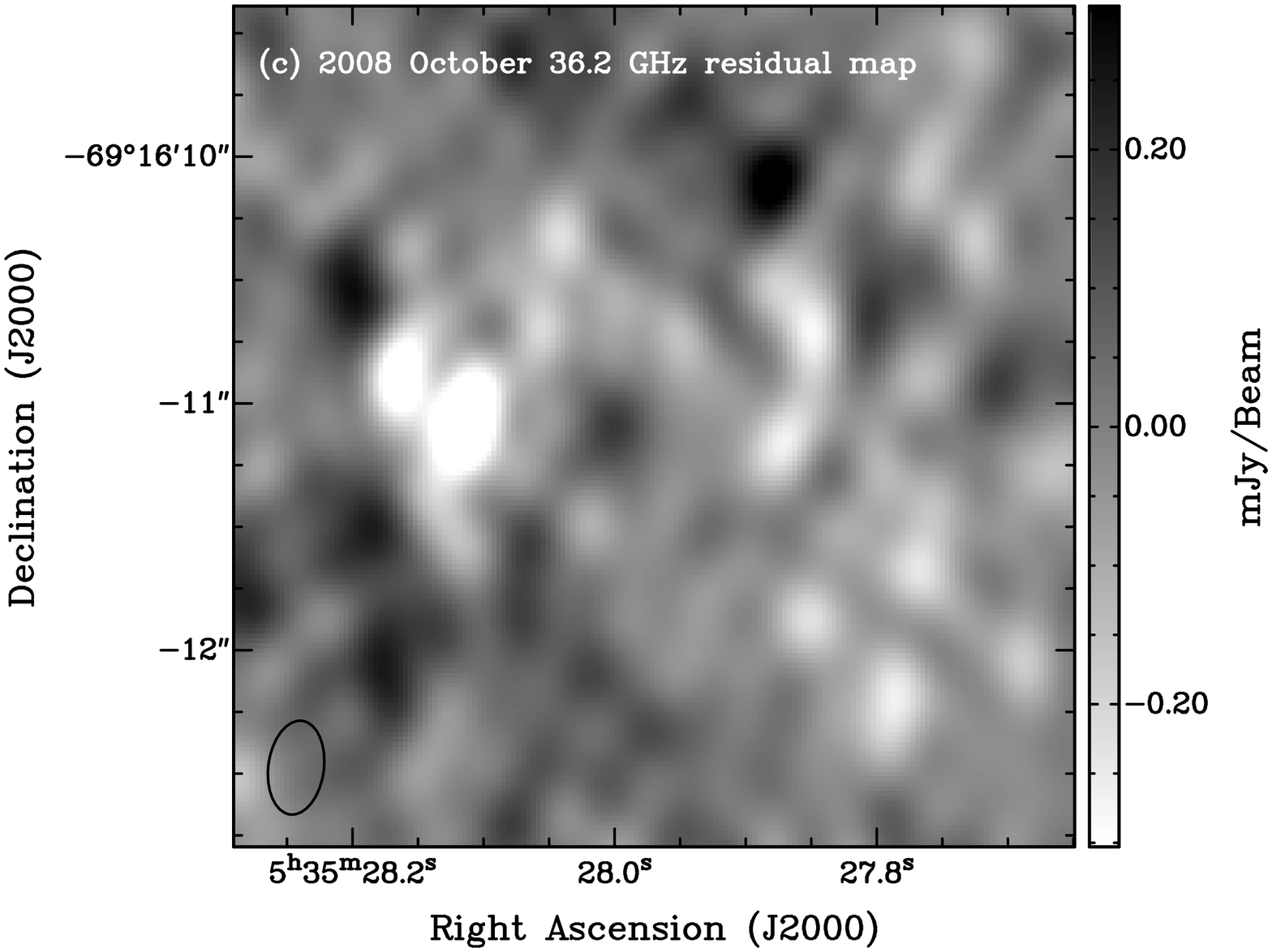}
\caption{(a) 2008 October 36.2~GHz image of SN1987A; compared with (b), the model from the Fourier-domain fitting \citep{YNg:2008p14427}; and (c) the imaged residual image between the image and model.  The real and model mages are displayed on a linear greyscale ranging from 0 to 4~mJy~beam$^{-1}$, and the residual image has a greyscale ranging $\pm$0.3~mJy~beam$^{-1}$. The rms noise of the residual image is estimated to be around 90 $\mu$Jy.}
\label{stephen_model}
\end{center}
\end{figure}

At 36.2~GHz, the model was subtracted from the data to produce a residual image as shown in Figure \ref{stephen_model}(c), alongside the imaged data and imaged model in Figures \ref{stephen_model}(a) and \ref{stephen_model}(b). Features present or absent in the fitted model appear as negative or positive components in the residual image. 
Within the remnant in the residual image, the highest positive peak is $+0.2 \pm 0.1$~mJy~beam$^{-1}$ and the lowest negative peak is $-0.2 \pm 0.1$~mJy~beam$^{-1}$ with an average difference of zero. The amplitude of the peak, the maximum difference between the real and model image, represents an upper limit on the flux density from a possible central source. A more quantitative measurement was made by adding an additional point source to the truncated torus model and fitting the model to the 2008 October uv-data at 36.2~GHz. The resulting flux density of the fitted point source in the model is $0.3 \pm 0.2$~mJy, which is consistent with the above estimate. It is also consistent with the 0.3~mJy upper limit at 9~GHz, estimated in \citet{YNg:2008p14427}. Possible reasons for this extra point source flux include: an energized population of accelerated electrons due to projected shell emission at high latitudes away from the equatorial ring, a source within the central region of the remnant, noise, or possibly phase errors in the 36.2~GHz image, Assuming 0.3~mJy as the flux density of a possible central source, we estimate a 36 GHz spectral luminosity of $1 \times 10^{14}$ Watts Hz$^{-1}$ at a distance of 51.4 kpc \citep{Panagia:2003p13411}. This value is significantly under-luminous when compared to some central sources of other known supernova remnants. \citet{Page:2007p13570} obtain a flux density of 322~Jy at 33~GHz and 299~Jy at 40~GHz from WMAP observations of the Crab Nebula. Assuming an interpolated flux density of 310 Jy for 36.2~GHz gives a luminosity for the Crab of $1.5 \times 10^{17}$ W Hz$^{-1}$ at a distance of 2 kpc \citep{Trimble:1973p13894}. Observations at 22~GHz of the young remnant of SN1986J \citep{Bietenholz:2008p13739} show a flux density of 4~mJy in a central region of flat spectral index. Extrapolated to 36.2~GHz, this corresponds to a flux density of 3~mJy or a luminosity of $4 \times 10^{19}$ W Hz$^{-1}$ at a distance of 10~Mpc. 

\subsection{Spectral index image}\label{specindexmap}

\begin{figure}[h]
\begin{center}
$\begin{array}{cc}
\includegraphics[width=7.65cm, angle=0]{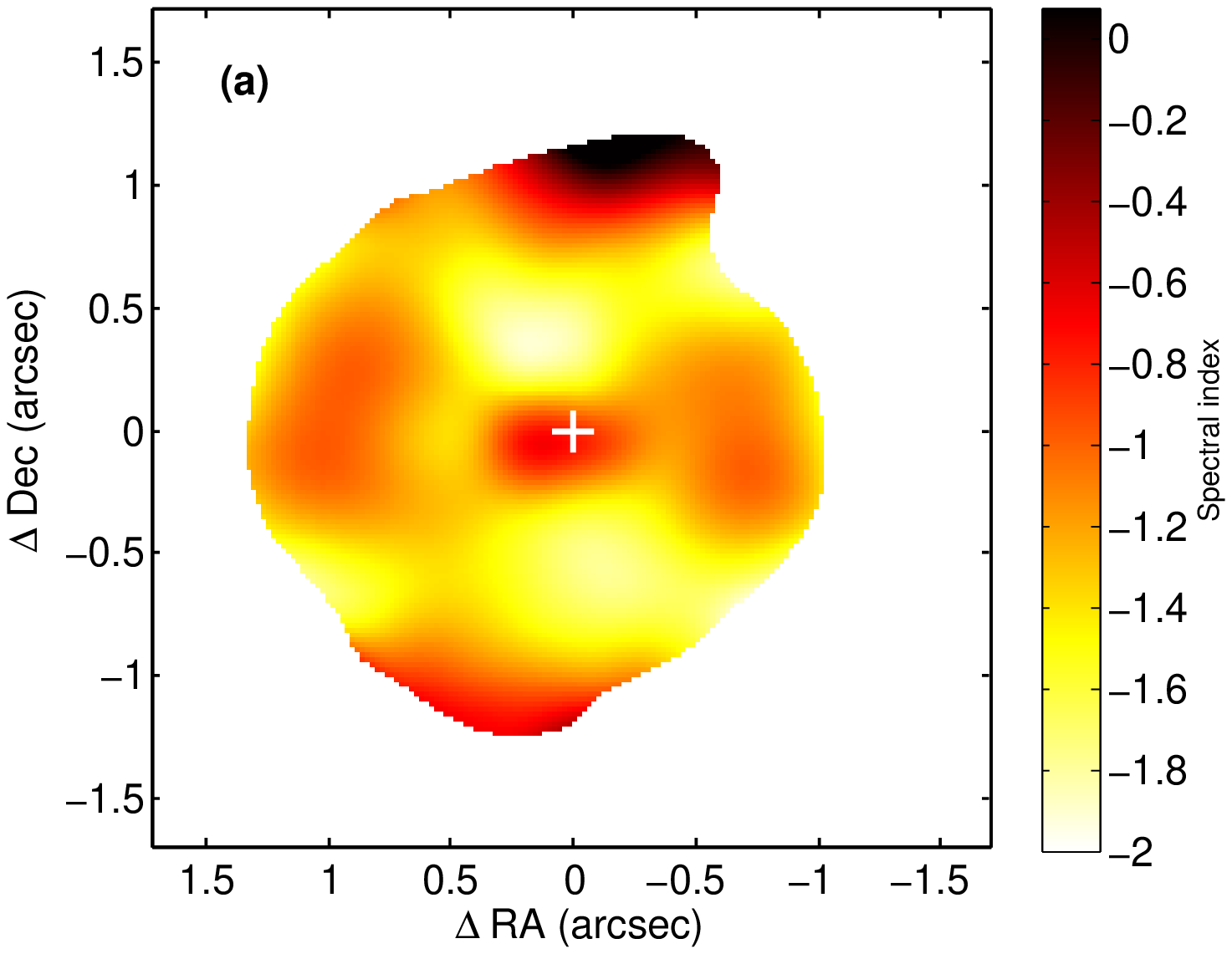}
\includegraphics[width=7.65cm, angle=0]{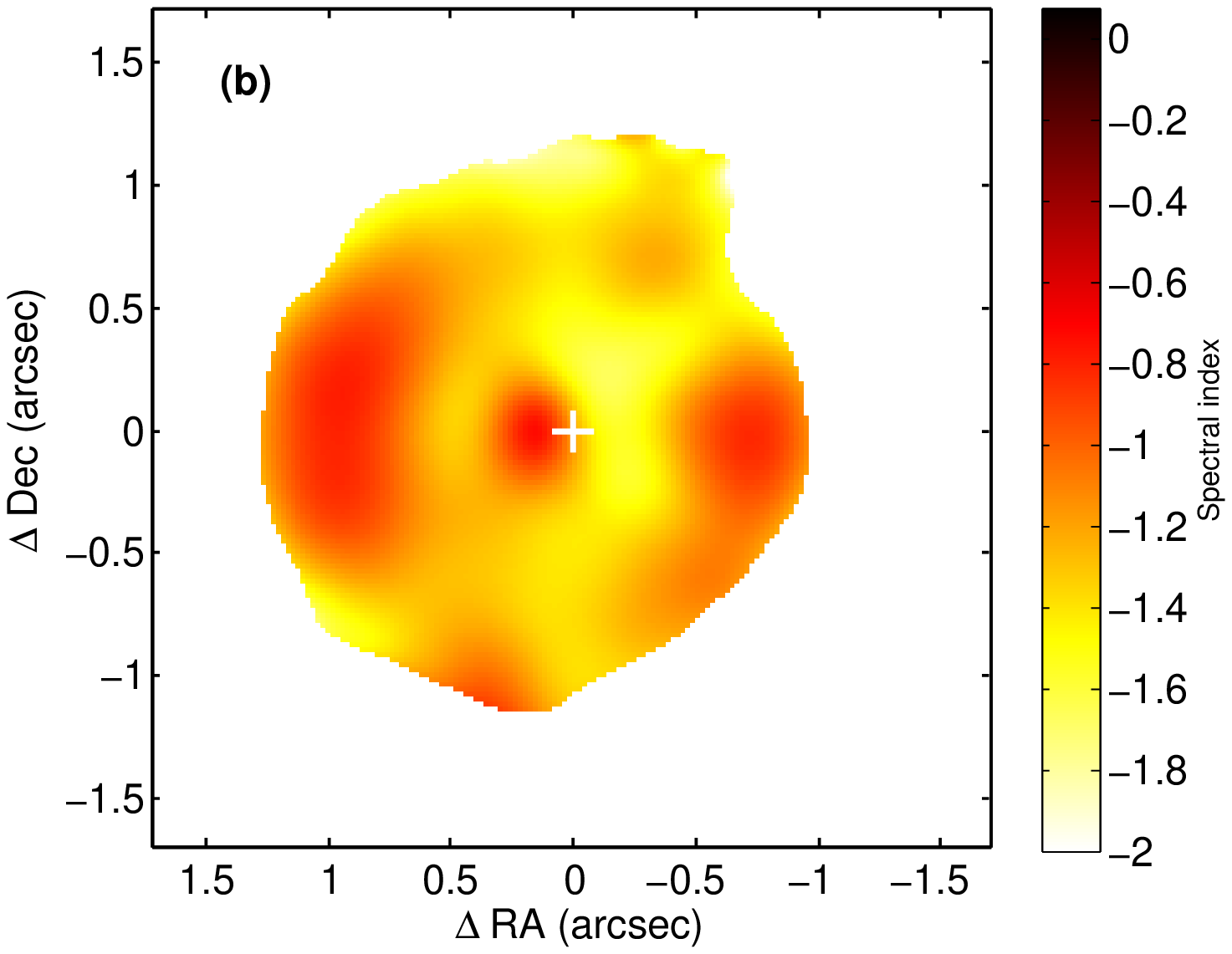}
\end{array}$
\caption{Spectral index images formed by comparing the super-resolved 2008 Oct 8.6~GHz and the re-imaged 2008 Oct 36.2~GHz images of SN1987A with the same MAXEN deconvolution algorithm and $0\farcs4$ restoring beam. (a) is the real spectral index image and (b) is the model spectral index image generated from the imaged Fourier models. The progenitor position is shown as a white cross in each image.}\label{spectralindexmaps}
\end{center}
\end{figure}

SN1987A has been `classified' as a Type IIP supernova with a progenitor mass of around 14.7 solar masses \citep{Nomoto:1989p4016}. We know from neutrino observations that a large number of neutrons were formed in the core collapse. However it is still uncertain if a neutron star survived the explosion. Searches for an associated pulsar have been unsuccessful so far \citep{Manchester:2007p3729}, although they only rule out the existence of a relatively high-luminosity pulsar. Another possible way to find evidence of a surviving neutron star is to look for a pulsar wind nebula. Pulsar wind nebulae (PWNe), are clouds of non-thermally radiating particles that are excited by an enclosed compact central source. PWNe have been detected in about ten percent of all known core collapse supernova remnants \citep{Kaspi:2002p14079}, with the Crab nebula and 3C58 being some of the the best-known examples. In the radio such nebulae are characterised by polarised synchrotron emission and typically have a flat spectral index of ranging between -0.3 and 0 \citep{Gaensler:2006p14115}. Detection of flat spectral index emission near the explosion site of SN1987A, as was the case with SN1986J \citep{Bietenholz:2004p13596}, would be a good indication that a neutron star has survived the explosion. A search for a possible central flat spectrum source was undertaken by generating a spectral index image using the 2008 Oct  data at 36.2~GHz and the 2008 October data at 8.6~GHz [discussed in Section \ref{obsreduc}]. 
As the synthesised beam for the 8.6~GHz data is 0\farcs4, the 36.2~GHz data were re-imaged with a 0\farcs4 restoring beam using the MAXEN algorithm instead of CLEAN for consistency with the 8.6~GHz image. Both the 8.6~GHz and the re-imaged 36.2~GHz images are shown in the middle row of Figure \ref{models_and_maps}. As a precaution against possible false spectral index information generated by the imaging process, we also produced a comparison spectral index image made from the Fourier models in the top row of Figure \ref{models_and_maps}. The models were inserted into the visibility data with a suitable position offset to locate it at the same angular distance from the phase center but on the opposite side of the phase center from the remnant. Following imaging and deconvolution, the resulting model images, shown in the bottom row of Figure \ref{models_and_maps}, have approximately the same noise, flux density, and restoring beam characteristics as their real counterparts. In Figure \ref{spectralindexmaps} we present the two spectral index images. Figure \ref{spectralindexmaps}(a) was formed by comparing the 2008 Oct 8.6~GHz and re-imaged 2008 Oct 36.2~GHz observations, and Figure \ref{spectralindexmaps}(b) was formed by comparing the 2008 Oct 8.6~GHz and 2008 Oct 36.2~GHz model images. The real spectral index image has been masked using the same mask used to determine the flux of the 2008 October CLEAN image at 36.2~GHz. A similar mask has also been applied to the model spectral index image.

Both spectral index images share the same broad characteristics, regions of spectral index, $\alpha \approx -0.8$ at the two lobes and regions of  steep spectral index, $\alpha \approx -1.5$ in the interior. There are two notable flat spectral index features in the real image. The feature at the northern rim is caused by the spurious north-south extensions mentioned in Section \ref{obsreduc}. The feature around 0\farcs2 east of the progenitor position, which has a maximum spectral index of $\alpha \approx -0.7 \pm 0.3$ and encompasses an area of approximately one beam (width $0.4\arcsec$). The position of the flattest spectral index in this feature is $05^{\rm{h}}35^{\rm{m}}27^{\rm{s}}.991,-69\arcdeg16\arcmin11\farcs13$ (J2000) with around 50 mas positional uncertainty. However a smaller flat spectral index feature is also present in the model counterpart. This feature is largely due to the imaging process rather than noise or differences between the models. Follow-up observations will be needed to determine if the feature in the real spectral index images corresponds to an actual source of emission in the  central region emission not fitted by our Fourier models.

 %At present, Diffusive Shock Acceleration (DSA) \citep{Duffy:1995p46,Melrose:2009p11879}, has the most potential for explaining particle acceleration at shock fronts. The theory predicts that the distribution of radio emission is a power law whose spectral index is related to the shock compression ratio r by $r=1-3/(2 \alpha)$. A spectral index of 0.08 cannot be accounted for by the theory, thus making a DSA related explanation for the flat spectral index feature implausible. This leaves non-DSA particle acceleration from either the shock or compact central source, non-thermal emission from the supernova ejecta, noise, or phase errors as possible explanations for the flat spectrum feature. Follow-up observations will be needed to distinguish which cause is more likely.

\subsection{Polarisation and magnetic field direction.}

Magnetic field direction in supernova remnants can be divided into three categories: tangled, radial and tangential. In the atlas of supernova remnants compiled by \citet{Milne:1987p11293} and \citet{Furst:2004p11325}, 27 out of the 70 remnants surveyed showed polarisation. The absence of polarisation in the presence of synchrotron emission indicates a tangled magnetic field and/or extreme rotation measures along the line of sight. Of those remnants which did exhibit polarisation, such as Cassiopeia A and Tycho's remnant, most under 10,000 years old had radial magnetic field directions across the shock front. Rayleigh-Taylor (R-T) instabilities have been implicated as the mechanism for the radial stretching of the magnetic field in young supernova remnants, however magnetohydrodynamical simulations have yet to produce R-T instabilities that reach across the forward shock \citep{Blondin:2001p12420}.  It is anticipated that higher shock compression ratios generated by the influence of shock-accelerated particles will enable the generation of radial magnetic fields that cross the shock \citep{Schure:2008p12137}. Older supernova remnants such as  Vela exhibit a tangential magnetic field, interpreted as the accumulation of an interstellar magnetic field at the ageing blast front.

A search was made for linear and circularly polarised emission in the remnant of SN1987A by constructing polarisation intensity images with natural weighting. No persistent linear polarisation in Stokes-Q was detected above a 3$\sigma$ threshold of 0.6 and 0.24~mJy~beam$^{-1}$ for the 2008 April and 2008 October observations. A similar technique was employed to search for circular polarisation in Stokes-V, however the same null result was obtained for a $3\sigma$ threshold of 0.6 and 0.24~mJy~beam$^{-1}$. Possible causes for the de-polarisation of the radio emission include a tangled magnetic field in the emitting region and/or extreme Faraday rotation measures on the order of $\approx 50000$ rad m$^{-2}$. 

\subsection{Radial profile and astrometry}\label{astrometry}

The optical position of the supernova at day 1278, as measured by a frame-tie to the Hipparcos and VLBI reference frames is $05^{\rm{h}}35^{\rm{m}}27^{\rm{s}}.968,-69\arcdeg16\arcmin11\farcs09$ (J2000), with an error of 30 milli-arcseconds (mas) in each coordinate \citep{Reynolds:1995p3810}. At day 7899, corresponding to 2008 Oct 9, this position will have moved due to the proper motion of the remnant. The proper motion of the LMC according to \citet{Kallivayalil:2006p14465} is $+2.03 \pm 0.08$ mas yr$^{-1}$ east and $+0.44 \pm 0.05$ mas yr$^{-1}$ north. The remnant also experiences a $0.3\pm 0.02$ mas yr$^{-1}$ motion north due to its estimated $65 \pm 5$~km s$^{-1}$ bulk rotation speed around the LMC \citep{Kim:1998p14519} at a distance of approximately 5 kpc from the LMC's kinematic center (assuming a distance of 50kpc to SN1987A). If corrected for both effects, then the progenitor should, at day 7899, be situated $37 \pm 1.5$ mas east and $13 \pm 1.3$ mas north of Reynold's position, assuming an LMC distance of 50 kpc. This predicted position corresponds to the coordinates of $05^{\rm{h}}35^{\rm{m}}27^{\rm{s}}.975,-69\arcdeg16\arcmin11\farcs08$ (J2000) with an uncertainty of 30 mas in each coordinate. We tested this prediction by fitting an ellipse to points that make up a ridgeline of the radio ring in the combined 2008 April-Oct 36.2~GHz image. Points along the ridgeline in the radio image were located by propagating rays away from the original progenitor position. The point of maximum flux density along each propagated ray was determined to be part of the ridgeline. We fitted an ellipse to both flux-weighted and non flux-weighted ridge line points and tabulated the fitted ellipsoids in Table \ref{ellipsefit}. We see that fitting an ellipse to the 36.2~GHz October data in the image plane produces a semi-major axis that is slightly smaller but consistent with the mean radius obtained by fitting a truncated-shell model to visibility data in the uv plane.

\begin{center}
\begin{table*}[ht]
\caption{Parameters of the flux-weighted and flux un-weighted elliptical fits.}\label{ellipsefit}
\begin{center}
\begin{minipage}{14.5cm}
\begin{center}
\begin{tabular}{lccccc}
\tableline \hline \small
Weighting & RA offset\footnotemark[1]   & Dec offset\footnotemark[1]    & Semi-major axis & Semi-minor axis  & Position angle\footnotemark[2]  \\
  &  (mas) &  (mas) &  (\arcsec) & (\arcsec) &  (\arcdeg) \\
\tableline Weighted & $+72  \pm 48 $ & $-37 \pm 46 $ & 0.79 & 0.68 & 74.7 \\
 None & $+79 \pm 48$  & $-32 \pm 46$ & 0.80 & 0.66 & 70.6 \\
\tableline 
\end{tabular} \newline
\end{center}
\textbf{Notes}. $^{1}$Offsets are the angular distances of the fitted ellipse centroids from the predicted position of  $05^{\rm{h}}35^{\rm{m}}27^{\rm{s}}.975,-69\arcdeg16\arcmin11\farcs08$ (J2000) \\
$^{2}$Position angle is defined North through East.
\end{minipage}
\end{center}
\end{table*}
\end{center}

\begin{figure*}[h]
\begin{center}
\includegraphics[width=16.5cm, angle=0]{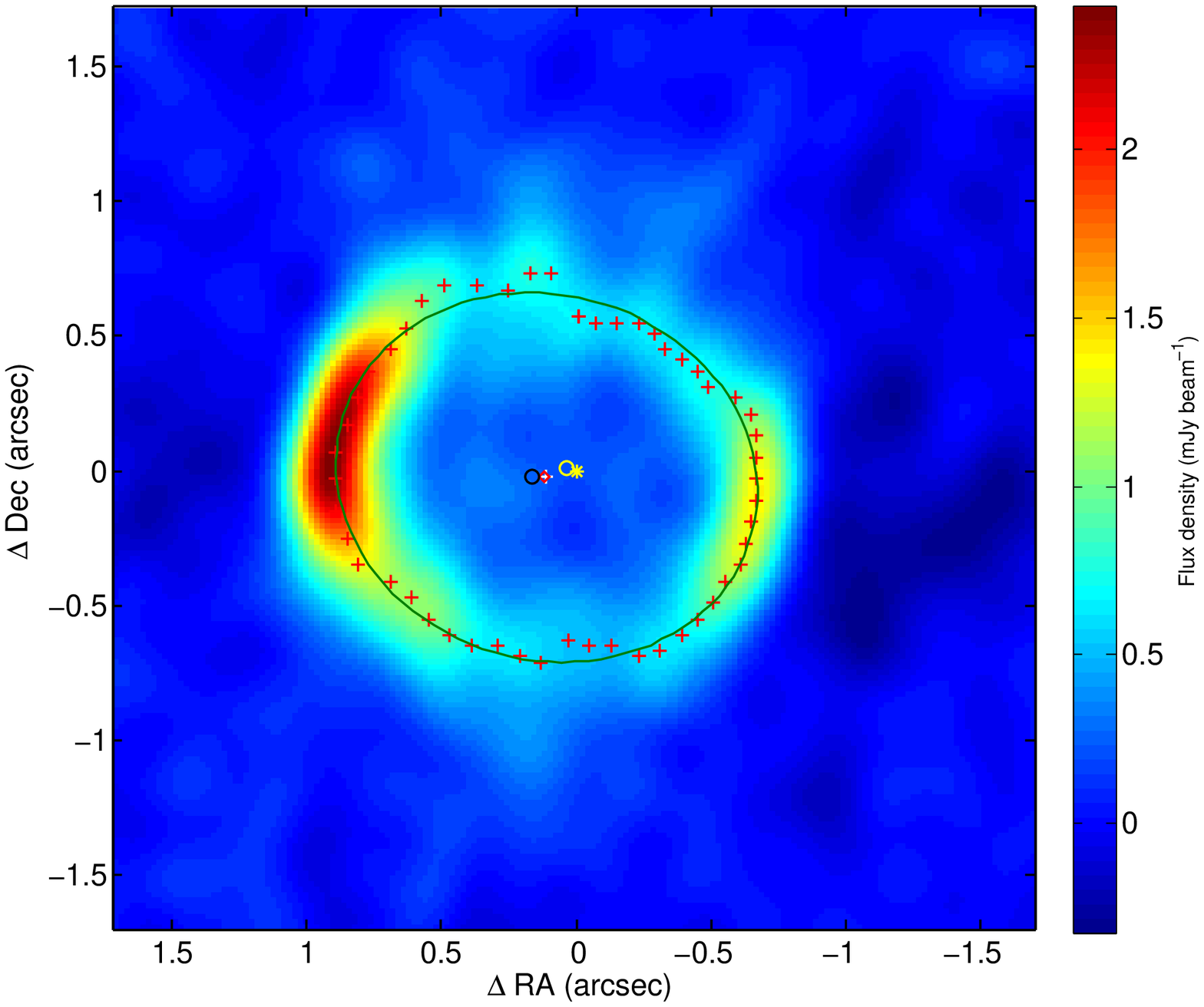}
\caption{Results from the ellipse fits to the morphology of the combined 2008 October and 2008 April 36.2~GHz observations. The green ellipse drawn atop the image is the flux-weighted ellipse fit to the red crosses returned by the ridge-following algorithm. The centroid of the green ellipse is shown as a white cross almost hidden under the red diamond which represents the centroid of the flux-unweighted ellipse fit. The yellow star is the measured optical position of the progenitor \citep{Reynolds:1995p3810}; this moves to the position of the small yellow circle when the proper motion of the remnant and the LMC is accounted for. The black circle represents the position of the flat spectral index feature.}\label{fittedellipse}
\end{center}
\end{figure*}

In Figure \ref{fittedellipse} the flux-weighted ellipse and ridgeline points are overlaid atop the combined 2008 Apr-Oct 36.2~GHz image along with the two fitted ellipse centroids and the original progenitor position. The flux-weighted ellipse centroid was located at $05^{\rm{h}}35^{\rm{m}}27^{\rm{s}}.987,-69\arcdeg16\arcmin11\farcs115$ (J2000) and the un-weighted centroid was located at $05^{\rm{h}}35^{\rm{m}}27^{\rm{s}}.990,-69\arcdeg16\arcmin11\farcs110$ (J2000). Because we have used the VLBI position of the phase calibrator 0530-727 from \citet{Reynolds:1995p3810}, the estimated positional uncertainty of the fitted centroids above is about 30 mas in each coordinate with an additional $\approx 15-17$ mas uncertainty in the fit. As shown in Figure \ref{fittedellipse}, the offset of the flux-weighted centroid is $72 \pm 48$ mas east and $37 \pm 45$ mas south of the predicted position, and the flux-unweighted centroid is $79 \pm 48$ mas east and $32 \pm 46$ mas south of the predicted position. 
Given that the positional uncertainty in the predicted position is about 30 mas, we conclude that the two fitted centroids and the predicted position all lie within $~1\sigma$ of each other. If a hydrodynamical effect following the explosion is the cause of the centroid offset from the predicted position, we estimate that the (day 7899) upper limit for any departure from symmetry for the eastern lobe, estimated as approximately twice the offset plus uncertainty, is at most 174 mas east and 64 mas south. The position of the possible flat spectral index feature in Figure \ref{spectralindexmaps}(a) was offset 124 mas east and 32 mas south of the predicted position, which places it within 60 mas of the fitted centroids.

\clearpage

\begin{figure}[h]
\begin{center}
\includegraphics[width=7.65cm, angle=0]{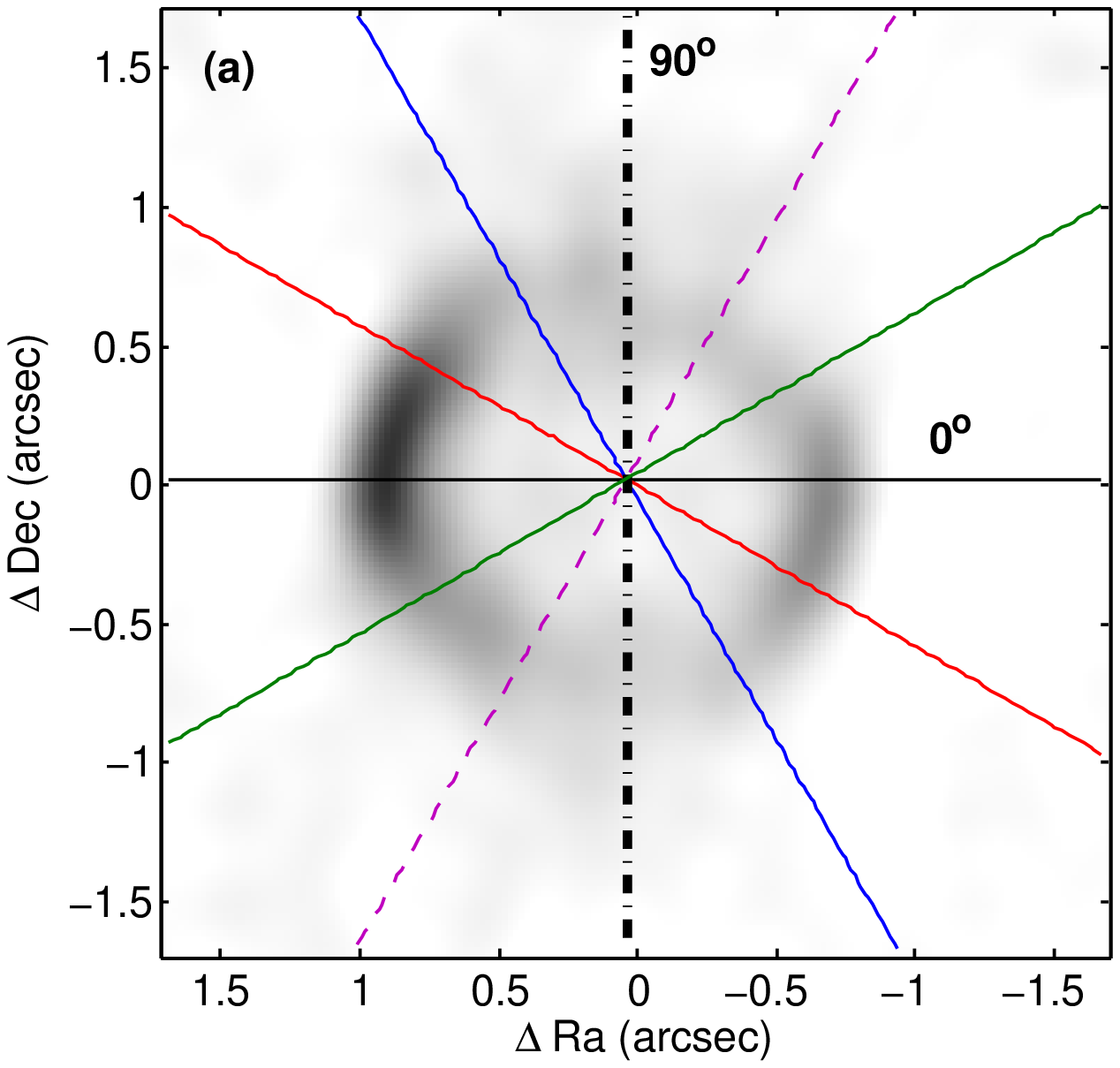}
\includegraphics[width=7.65cm, angle=0]{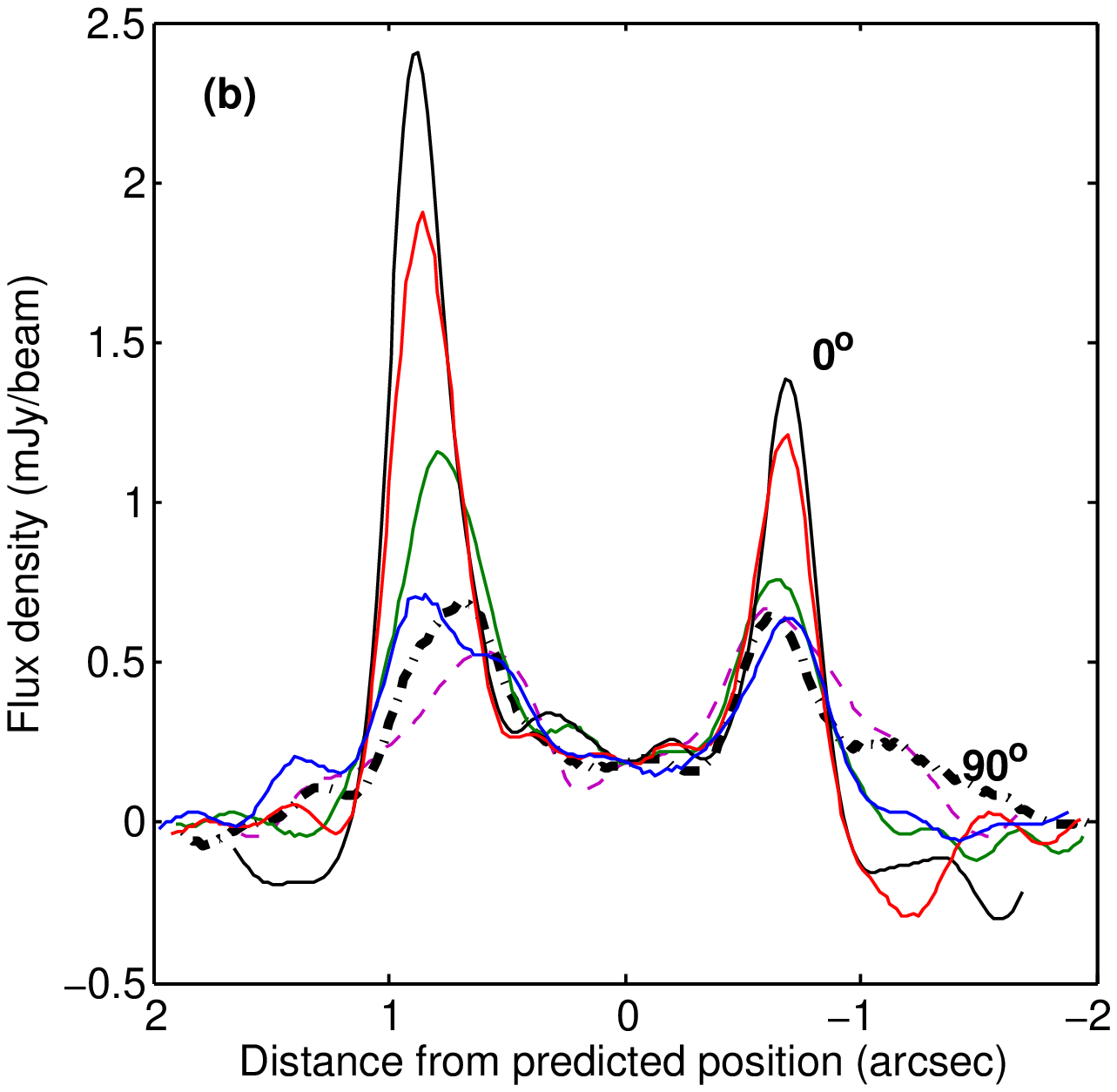}
\caption{Radial profiles overlaid on the combined 2008 Apr and 2008 October 36.2~GHz image. Radial lines are propagated away from the predicted progenitor position in (a) and their radial distances are measured in (b). The positive distances measured in (b) correspond to rays propagating through the left half of (a).}\label{radial_lines}
\end{center}
\end{figure}

Radial profiles of the remnant were made by propagating rays away from the averaged position of the ellipse centroid [obtained from the ridge-finding algorithm] and plotting the resulting profiles, shown in Figure \ref{radial_lines}. In particular, note that the flux density within the projected radio ring on the image plane is always greater than $6\sigma$ above the rms noise of 0.06~mJy~beam$^{-1}$, which indicates that emission is being produced along the line of sight, either from high latitude regions of the expanding shell, or from within the central region of the nebula itself.  
 
\subsection{Comparison with observations at other wavelengths}

\begin{figure*}[h]
\begin{center}$
\begin{array}{cc}
\includegraphics[width=6.5cm, angle=270]{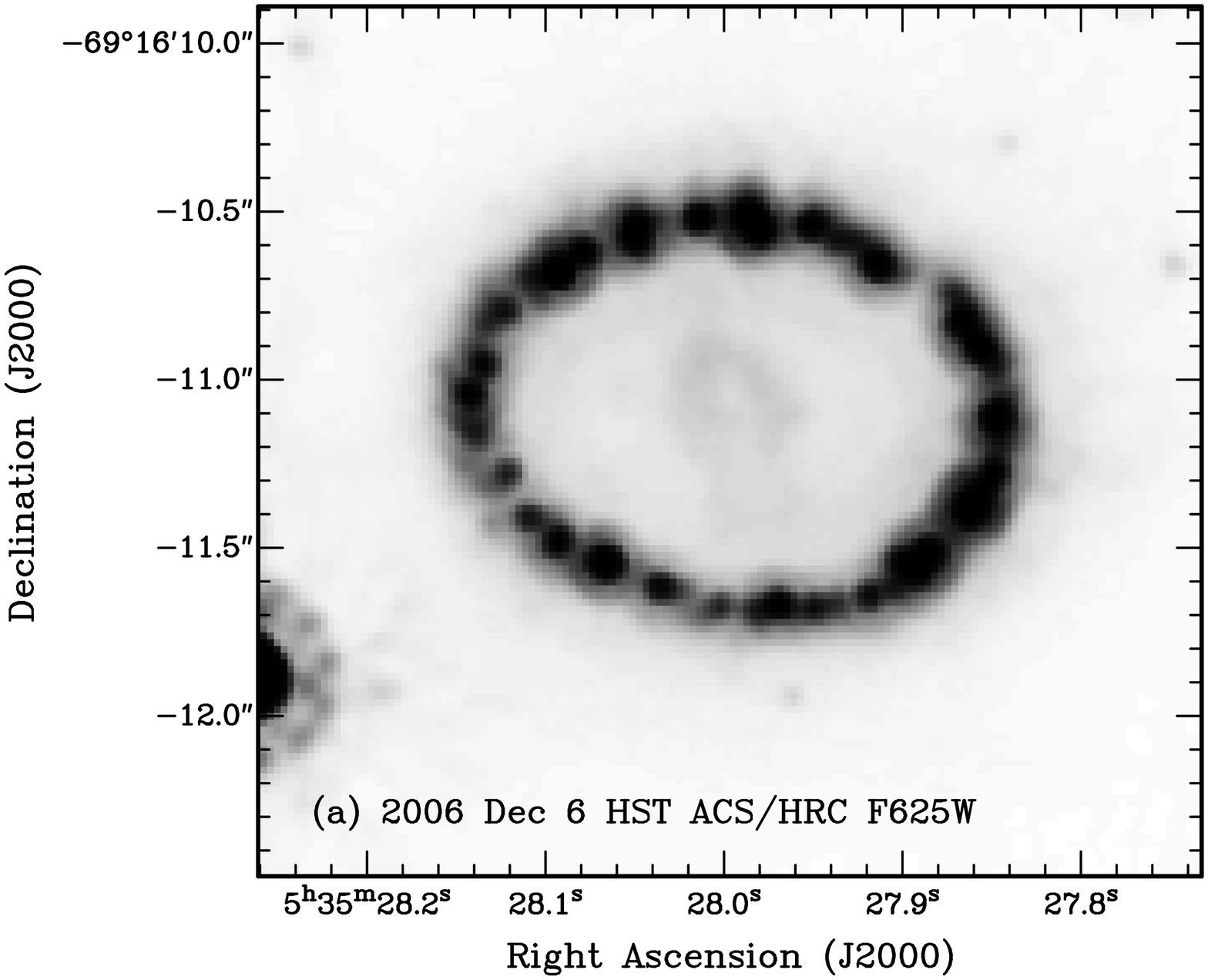} &
\includegraphics[width=6.5cm, angle=270]{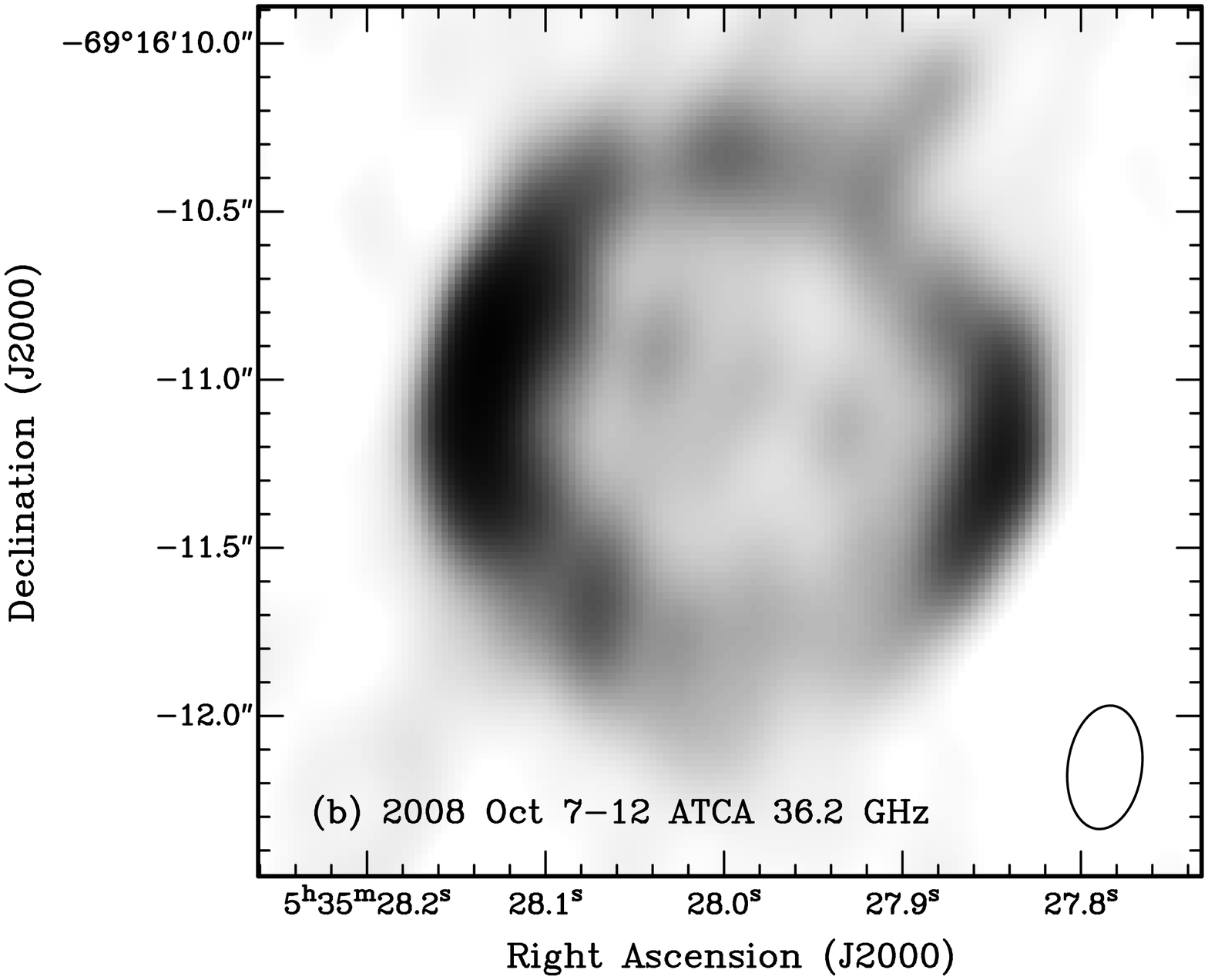} \\
\includegraphics[width=6.5cm, angle=270]{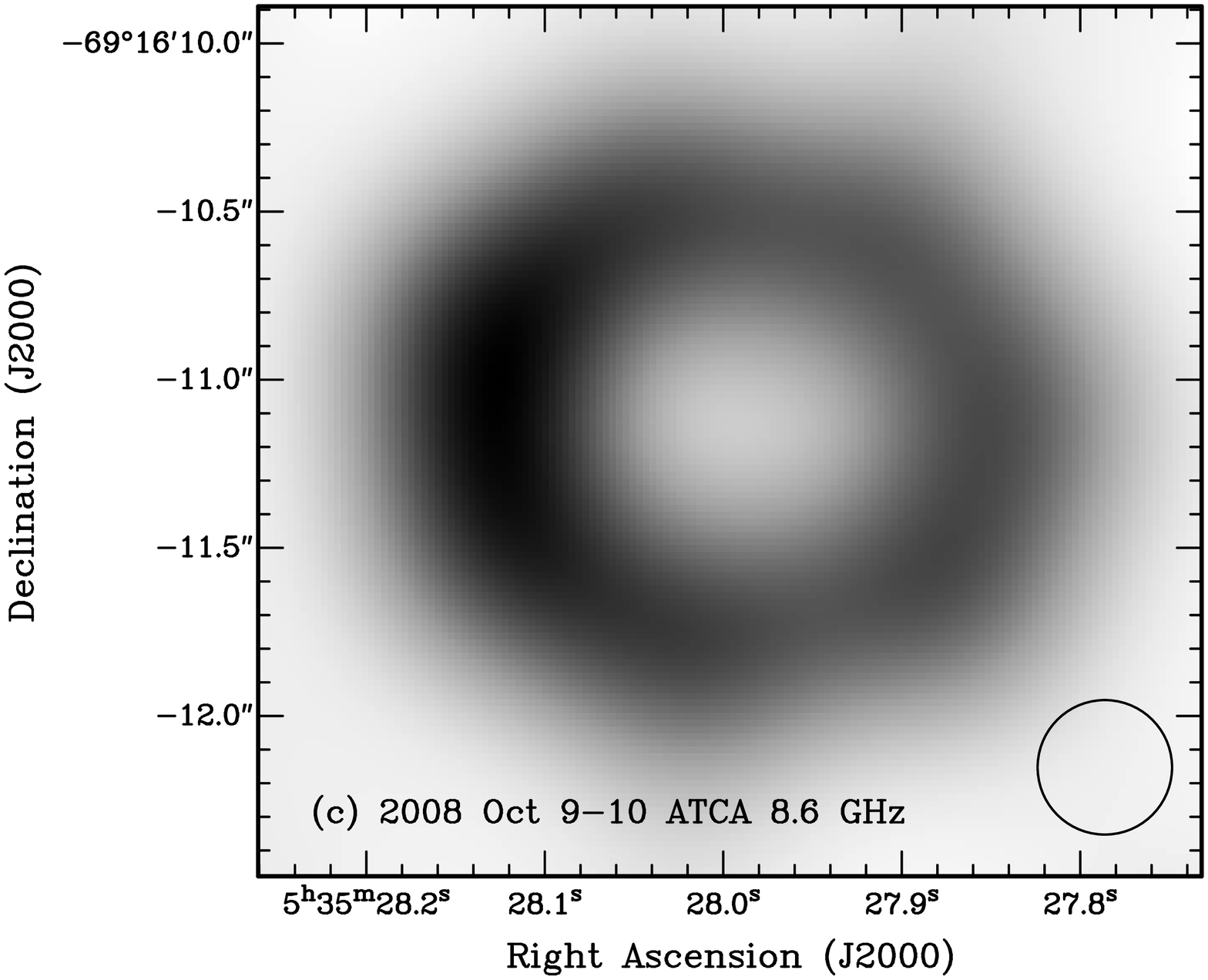} &
\includegraphics[width=6.5cm, angle=270]{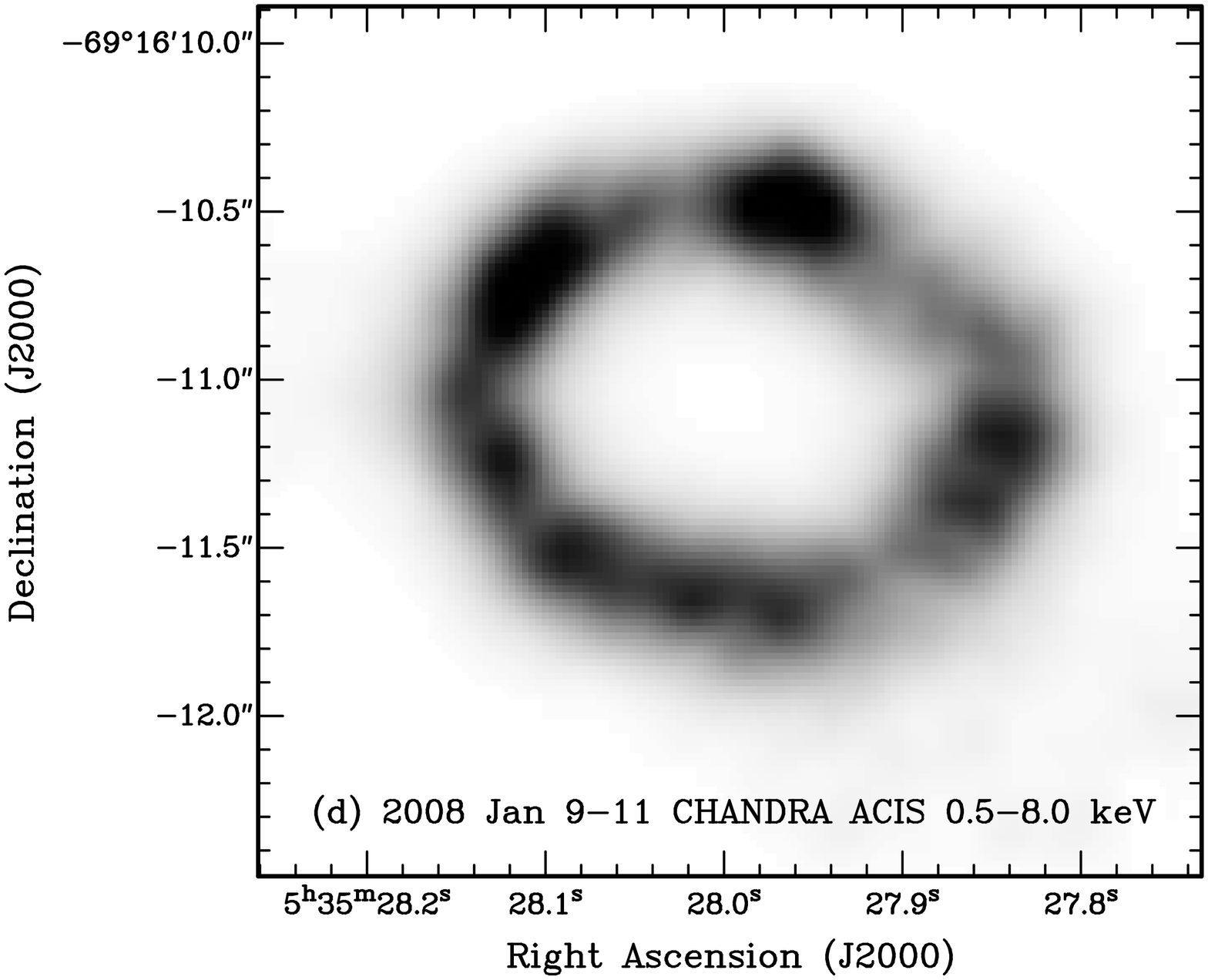} \\
\end{array}$
\end{center}
\caption{SN1987A at different wavelengths, (a) is the 2006 Dec 6 HST image, (b) is the 2008 Oct 7-12 ATCA 36.2~GHz continuum image at robust=0.5 weighting, (c) is the 2008 Oct 9-10 ATCA 8.6~GHz super-resolved image at uniform weighting  and (d) is the 2008 Jan 9-11 Chandra 0.5-8.0 keV image. The 8.6~GHz and 36.2~GHz images have a greyscale range of 0-12~mJy~beam$^{-1}$ and 0-4~mJy~beam$^{-1}$. Images (a) and (d) have had their coordinates shifted to coincide with the radio images as in Figure \ref{rgb_overlay}. The linear greyscale on each image has been stretched to highlight intrinsic low brightness features. }
\label{four_way_comparison}
\end{figure*}

In order to see how the 2008 Oct 8.6~GHz and 2008 Oct 36.2~GHz images compared with observations at other wavelengths we obtained optical Hubble Space Telescope (HST) and X-ray Chandra images from their respective public archives. The HST image was taken on 2006 Dec 6 using filter F625W in ACS/HRC configuration (dataset id\# J9OT07070). The Chandra X-ray image was a 0.5-8.0 keV ACIS image from an observation taken on 2008 July 9-11. The X-ray image was reprocessed following the procedure described in \citet{Park:2006p44}. We removed the pixel randomization and applied the `subpixel resolution' technique \citep{Tsunemi:2001p8231} to further improve the effective spatial resolution of the image. It was then deconvolved with the Lucy-Richardson algorithm \citep{richardson:1972,Lucy:1974p8526} using a monochromatic PSF of 1.5 keV (2.5 keV for the hard band) generated by the Chandra Ray Tracer and MARX. The results were smoothed by a Gaussian of 0\farcs1 FWHM. Shown in Figure \ref{four_way_comparison} is a comparison between the HST, 36.2~GHz ATCA, 8.6~GHz ATCA, and Chandra X-ray images. The coordinates of the HST and Chandra images have been shifted such that the optical and X-ray rings are centered on the 2008 Oct 36.2~GHz radio ring with an uncertainty of roughly 50 mas in each coordinate; it also implies the central optical nebulosity aligns with the  flat spectrum radio component. However this step precludes us from making further judgements about an asymmetry in the radio expansion with respect to the optical ring. One thing that is interesting about this plot, however, is that the radio images are more circular than the optical and X-ray. If we apply the ridge-following algorithm to the flux-unweighted optical ring in the HST image, the width of the fitted ellipse semi-major axis is $0\farcs78$ and the semi-minor axis is $0\farcs57$ with about 40 mas uncertainty in each axis. Compared to the radio ring of size $0\farcs79 \times 0\farcs68$ with similar uncertainty, the optical ring is slightly more elliptical. The difference in ellipticity can be explained by the difference in location at which the different types of emission are generated. Non thermal radio emission is produced as the result of stochastic acceleration of particles at a shock front, and is primarily dependent on density for particle injection into the shock. Thermal X-rays are produced by material heated by the shock and is dependent on both density and temperature of the heated material. Hence we would expect radio emission to trace shock fronts to higher latitudes above the equatorial plane and X-ray and optical images to trace the higher density of heated material in the equatorial ring.

\clearpage

\begin{figure*}[h] 
\begin{center}
\includegraphics[width=13cm, angle=0]{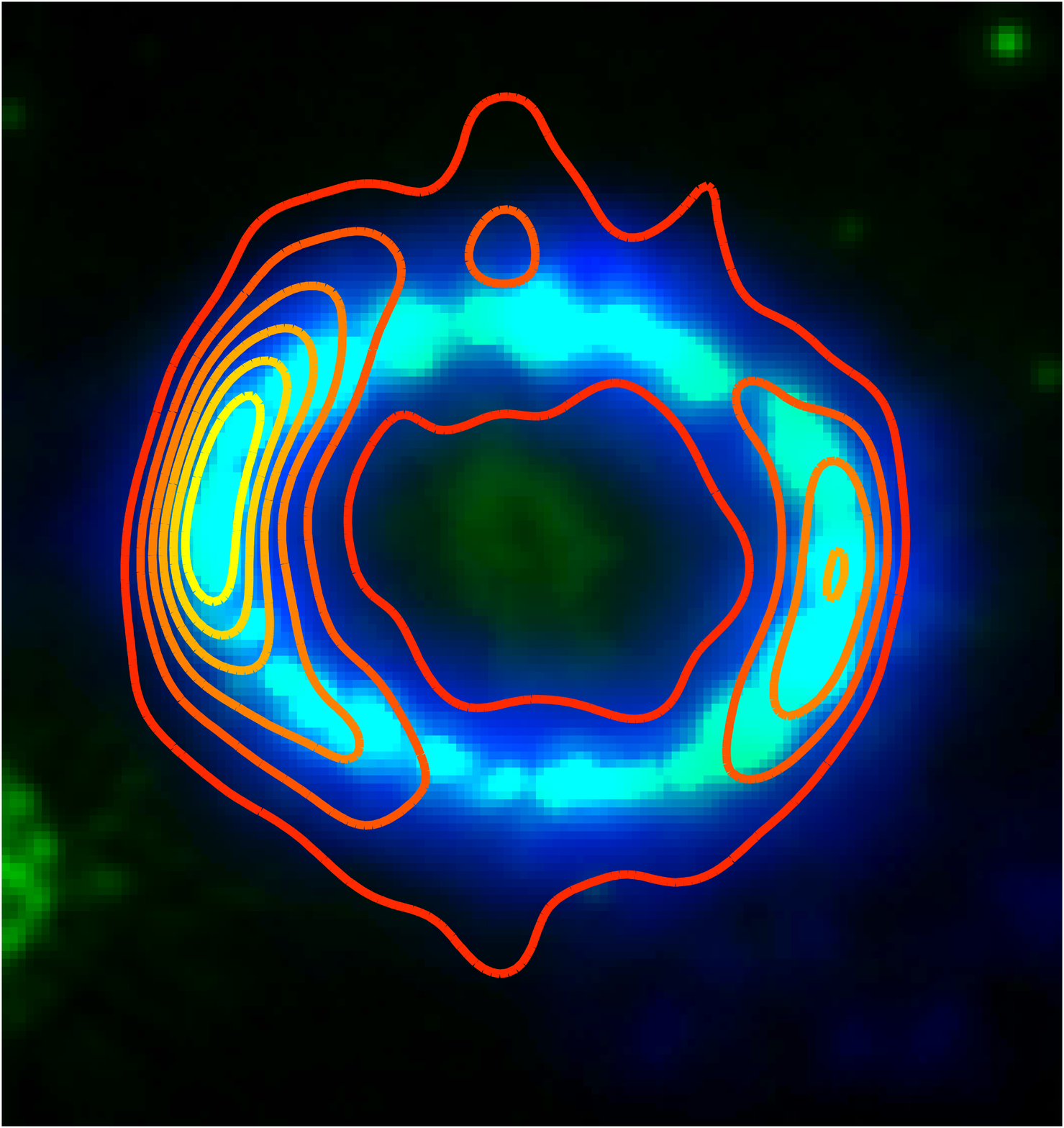}
\caption{Overlay of the combined HST 2006 Dec 6 optical (green), Chandra 2008 Jan 9-11 X-ray (blue), and ATCA 2008 Oct 36.2~GHz radio images (orange-yellow contours) formed by shifting the optical and X-ray coordinate systems to center on the radio ring from the 2008 Oct 36.2~GHz radio image at robust=0.5 weighting. Radio contours are at 14 (orange), 30, 40, 60, 70 and 85\% (yellow) of the maximum at 2.4~mJy~beam$^{-1}$. The outermost contour and the contour within the optical ring are at the same 14\% level.}\label{rgb_overlay}
\end{center}
\end{figure*}

In figure \ref{rgb_overlay} the HST and Chandra images have been overlaid with the radio contours from the 36.2~GHz radio image. From the peaks in the radio contours it is evident that the forward shock is only slightly larger than the 0\farcs8 optical ring. The truncated-shell models fitted to the 2008 October data give a more accurate shell radius of $0\farcs85 \pm 0\farcs06$ for the 36~GHz data which is larger but still consistent. However the radius measured at 8.6~GHz is $0\farcs892\pm 0\farcs002$, which suggests that the forward shock has almost completed its transit of the optical ring.

\clearpage

\section{Conclusions}

In this paper we have presented results from the 36.2~GHz extension of the monitoring observations of SN1987A performed in April and October 2008. The October observation was conducted under conditions of good atmospheric stability. A flux density of $27 \pm 6$~mJy was obtained which allows us to derive a spectral index of $\alpha \approx -0.83$ when fitted along with flux densities from the lower frequency monitoring observations. The Fourier modeling technique of \citet{YNg:2008p14427} was employed to produce a model fit of the torus that is consistent with the 8.6~GHz image. A comparison of the 36.2~GHz image with the 8.6~GHz image at similar resolution was used to obtain a spectral index image that showed a region of flatter spectral index about 0\farcs2 east of the progenitor. Simulations suggest a component of this may be an artifact of the imaging process. However  follow-up observations are suggested to determine if the remaining component is real central region emission not fitted by our Fourier models. Higher sensitivity observations with the newly-commissioned Compact Array Broadband Backend (CABB) will help further clarify this feature. An elliptical fit to the 36.2~GHz radio morphology shows that the centroid of the remnant is within positional accuracy of the progenitor's location when corrected for the proper motion of the LMC and the motion of SN1987A within the LMC. The fit is also located within positional uncertainty of the flat spectral index feature. Comparisons between radio and X-ray images show that the remnant is more spherical in the radio, which can be explained by two different emission mechanisms at work. From the Fourier models and comparisons to the optical emission we conclude that the forward shock in the remnant is completing its transit of the optical ring. 

\acknowledgments{}

We would like to thank Bjorn Emonts and Robin Wark their assistance with our observations, Andre Fletcher for useful discussions on radio reduction, Brian Reville and John Kirk for their insightful comments on particle acceleration at shock fronts.  This research has made use of data obtained from the Chandra Data Archive. Some of the data presented in this paper were obtained from
the Multimission Archive at the Space Telescope Science Institute
(MAST). STScI is operated by the Association of Universities for Research
in Astronomy, Inc., under NASA contract NAS5-26555. Support for MAST for
non-HST data is provided by the NASA Office of Space Science via grant
NAG5-7584 and by other grants and contracts. C.-Y.N. and B.M.G. acknowledge the support of the Australian
Research Council through grant FF0561298.


%% in parentheses, after the keyword, but they will not be verified.

%{\it Facilities:} \facility{Nickel}, \facility{ (STIS)}, \facility{CXO (ASIS)}.




\begin{thebibliography}{0}
\expandafter\ifx\csname natexlab\endcsname\relax\def\natexlab#1{#1}\fi

\end{thebibliography}


\begin{thebibliography}{55}
\expandafter\ifx\csname natexlab\endcsname\relax\def\natexlab#1{#1}\fi

\bibitem[{Ball {et~al.}(2001) Ball, Crawford, Hunstead, Klamer, \&
  McIntyre}]{Ball:2001p6811}
Ball, L., Crawford, D.~F., Hunstead, R.~W., Klamer, I., \& McIntyre, V.~J.
  2001, \apj, 549, 599, (c) 2001: The American
  Astronomical Society

\bibitem[{Bietenholz \& Bartel(2008)}]{Bietenholz:2008p13739}
Bietenholz, M.~F., \& Bartel, N. 2008, Advances in Space Research, 41, 424

\bibitem[{Bietenholz {et~al.}(2004)Bietenholz, Bartel, \&
  Rupen}]{Bietenholz:2004p13596}
Bietenholz, M.~F., Bartel, N., \& Rupen, M.~P. 2004, Science, 304, 1947, (c)
  2004: Science

\bibitem[{Bionta {et~al.}(1987)Bionta, Blewitt, Bratton, Caspere, \&
  Ciocio}]{Bionta:1987p13260}
Bionta, R.~M., Blewitt, G., Bratton, C.~B., Caspere, D., \& Ciocio, A. 1987,
  Physical Review Letters (ISSN 0031-9007), 58, 1494

\bibitem[{Blondin \& Ellison(2001)}]{Blondin:2001p12420}
Blondin, J.~M., \& Ellison, D.~C. 2001, \apj, 560, 244,
  (c) 2001: The American Astronomical Society

\bibitem[{Blondin \& Lundqvist(1993)}]{Blondin:1993p14977}
Blondin, J.~M., \& Lundqvist, P. 1993, \apj, 405, 337

\bibitem[{Briggs(1995)}]{Briggs:1995p8023}
Briggs, D.~S. 1995, American Astronomical Society, 187, 1444, (c) 1995:
  American Astronomical Society

\bibitem[{Brooks(2007)}]{atnf-7mm}
Brooks, K. 2007, ATCA@7mm, Sydney, NSW: CSIRO (ATNF), http://www.atnf.csiro.au/observers/docs/7mm/

\bibitem[{Chevalier \& Dwarkadas(1995)}]{Chevalier:1995p4450}
Chevalier, R.~A., \& Dwarkadas, V.~V. 1995, \apjl, 452, L45, (c) 1995: The American Astronomical Society

\bibitem[{Clark(1980)}]{Clark:1980p9245}
Clark, B.~G. 1980, Astronomy and Astrophysics, 89, 377, a{\&}AA ID.
  AAA028.021.016

\bibitem[{Duffy {et~al.}(1995)Duffy, Ball, \& Kirk}]{Duffy:1995p46}
Duffy, P., Ball, L., \& Kirk, J.~G. 1995, \apj, 447, 364

\bibitem[{F{\"u}rst \& Reich(2004)}]{Furst:2004p11325}
F{\"u}rst, E., \& Reich, W. 2004, The Magnetized Interstellar Medium, 141

\bibitem[{Gaensler {et~al.}(1997)Gaensler, Manchester, Staveley-Smith,
  Tzioumis, Reynolds, \& Kesteven}]{Gaensler:1997p7998}
Gaensler, B.~M., Manchester, R.~N., Staveley-Smith, L., Tzioumis, A.~K.,
  Reynolds, J.~E., \& Kesteven, M.~J. 1997, \apj, 479,
  845, (c) 1997: The American Astronomical Society

\bibitem[{Gaensler \& Slane(2006)}]{Gaensler:2006p14115}
Gaensler, B.~M., \& Slane, P.~O. 2006, Annual Review of Astronomy {\&}
  Astrophysics, 44, 17

\bibitem[{Gaensler {et~al.}(2007)Gaensler, Staveley-Smith, Manchester,
  Kesteven, Ball, \& Tzioumis}]{Gaensler:2007p42}
Gaensler, B.~M., Staveley-Smith, L., Manchester, R.~N., Kesteven, M.~J., Ball,
  L., \& Tzioumis, A.~K. 2007, SUPERNOVA 1987A: 20 YEARS AFTER: Supernovae and
  Gamma-Ray Bursters. AIP Conference Proceedings, 937, 86

\bibitem[{Hirata {et~al.}(1987)Hirata, Kajita, Koshiba, Nakahata, \&
  Oyama}]{Hirata:1987p4127}
Hirata, K., Kajita, T., Koshiba, M., Nakahata, M., \& Oyama, Y. 1987, Physical
  Review Letters (ISSN 0031-9007), 58, 1490

\bibitem[{H{\"o}gbom(1974)}]{Hogbom:1974p9252}
H{\"o}gbom, J.~A. 1974, Astronomy and Astrophysics Supplement, 15, 417, a{\&}AA
  ID. AAA011.033.031

\bibitem[{Janka(1997)}]{Janka:1997p13373}
Janka, H.~T. 1997, eprint arXiv, 9013

\bibitem[{Jauncey {et~al.}(1988)Jauncey, Kemball, Bartel, Shapiro, Whitney,
  Rogers, Preston, \& Clark}]{Jauncey:1988p3319}
Jauncey, D.~L., Kemball, A., Bartel, N., Shapiro, I.~I., Whitney, A.~R.,
  Rogers, A. E.~E., Preston, R.~A., \& Clark, T.~A. 1988, Nature (ISSN
  0028-0836), 334, 412

\bibitem[{Kallivayalil {et~al.}(2006)Kallivayalil, van~der Marel, Alcock,
  Axelrod, Cook, Drake, \& Geha}]{Kallivayalil:2006p14465}
Kallivayalil, N., van~der Marel, R.~P., Alcock, C., Axelrod, T., Cook, K.~H.,
  Drake, A.~J., \& Geha, M. 2006, \apj, 638, 772, (c)
  2006: The American Astronomical Society

\bibitem[{Kaspi \& Helfand(2002)}]{Kaspi:2002p14079}
Kaspi, V.~M., \& Helfand, D.~J. 2002, Neutron Stars in Supernova Remnants, 271,
  3

\bibitem[{Kim {et~al.}(1998)Kim, Staveley-Smith, Dopita, Freeman, Sault,
  Kesteven, \& McConnell}]{Kim:1998p14519}
Kim, S., Staveley-Smith, L., Dopita, M.~A., Freeman, K.~C., Sault, R.~J.,
  Kesteven, M.~J., \& McConnell, D. 1998, \apj, 503,
  674, (c) 1998: The American Astronomical Society

\bibitem[{Lucy(1974)}]{Lucy:1974p8526}
Lucy, L.~B. 1974, Astronomical Journal, 79, 745, a{\&}AA ID. AAA011.031.061

\bibitem[{Luo \& McCray(1991)}]{Luo:1991p372}
Luo, D., \& McCray, R. 1991, \apj, 379, 659

\bibitem[{Manchester(2007)}]{Manchester:2007p3729}
Manchester, R.~N. 2007, SUPERNOVA 1987A: 20 YEARS AFTER: Supernovae and
  Gamma-Ray Bursters. AIP Conference Proceedings, 937, 134, (c) 2007: American
  Institute of Physics

\bibitem[{Manchester {et~al.}(2005)Manchester, Gaensler, Staveley-Smith,
  Kesteven, \& Tzioumis}]{Manchester:2005p7378}
Manchester, R.~N., Gaensler, B.~M., Staveley-Smith, L., Kesteven, M.~J., \&
  Tzioumis, A.~K. 2005, \apj, 628, L131, (c) 2005: The
  American Astronomical Society

\bibitem[{Manchester {et~al.}(2002)Manchester, Gaensler, Wheaton,
  Staveley-Smith, Tzioumis, Bizunok, Kesteven, \& Reynolds}]{Manchester:2002p1}
Manchester, R.~N., Gaensler, B.~M., Wheaton, V.~C., Staveley-Smith, L.,
  Tzioumis, A.~K., Bizunok, N.~S., Kesteven, M.~J., \& Reynolds, J.~E. 2002,
 \pasa, 19, 207

\bibitem[{Martin \& Arnett(1995)}]{Martin:1995p11941}
Martin, C.~L., \& Arnett, D. 1995, \apj, 447, 378

\bibitem[{McMaster \& Biretta(2008)}]{biretta:2008p10404}
McMaster, M., \& Biretta, J. 2008, WFPC2 Instrument Handbook, 10th edn. (3700
  San Martin Drive Baltimore, Maryland 21218: Space Telescope Science
  Institute), 338

\bibitem[{Melrose(2009)}]{Melrose:2009p11879}
Melrose, D.~B. 2009, eprint arXiv, 0902, 1803

\bibitem[{Michael {et~al.}(2003)Michael, McCray, Chevalier, Filippenko,
  Lundqvist, Challis, Sugerman, Lawrence, Pun, Garnavich, Kirshner, Crotts,
  Fransson, Li, Panagia, Phillips, Schmidt, Sonneborn, Suntzeff, Wang, \&
  Wheeler}]{Michael:2003p28}
Michael, E., {et~al.} 2003, \apj, 593, 809

\bibitem[{Milne(1987)}]{Milne:1987p11293}
Milne, D.~K. 1987, Joint USSR-Australia Shklovskii Memorial Symposium on
  Supernova Remnants and Pulsars, 40, 771

\bibitem[{Murphy {et~al.}(2008)Murphy, Gaensler, \&
  Chatterjee}]{Murphy:2008p14040}
Murphy, T., Gaensler, B.~M., \& Chatterjee, S. 2008, Monthly Notices of the
  Royal Astronomical Society: Letters, 389, L23, (c) Journal compilation
  {\copyright} 2008 RAS

\bibitem[{Ng {et~al.}(2008)Ng, Gaensler, Staveley-Smith, Manchester, Kesteven,
  Ball, \& Tzioumis}]{YNg:2008p14427}
Ng, C.~Y., Gaensler, B.~M., Staveley-Smith, L., Manchester, R.~N., Kesteven,
  M.~J., Ball, L., \& Tzioumis, A.~K. 2008, arXiv, astro-ph

\bibitem[{Nomoto {et~al.}(1989)Nomoto, Hashimoto, Shigeyama, Kumagai, Yamaoka,
  \& Sato}]{Nomoto:1989p4016}
Nomoto, K., Hashimoto, M., Shigeyama, T., Kumagai, S., Yamaoka, H., \& Sato, H.
  1989, Big Bang, 495

\bibitem[{Page {et~al.}(2007)Page, Hinshaw, Komatsu, Nolta, Spergel, Bennett,
  Barnes, Bean, Dor{\'e}, Dunkley, Halpern, Hill, Jarosik, Kogut, Limon, Meyer,
  Odegard, Peiris, Tucker, Verde, Weiland, Wollack, \&
  Wright}]{Page:2007p13570}
Page, L., {et~al.} 2007, \apjs, 170, 335,
  (c) 2007: The American Astronomical Society

\bibitem[{Panagia(2003)}]{Panagia:2003p13411}
Panagia, N. 2003, eprint arXiv

\bibitem[{Park {et~al.}(2006)Park, Zhekov, Burrows, Racusin, McCray, \&
  Borkowski}]{Park:2006p44}
Park, S., Zhekov, S.~A., Burrows, D.~N., Racusin, J.~L., McCray, R., \&
  Borkowski, K.~J. 2006, In: Proceedings of the "The X-ray Universe 2005", 604,
  335

\bibitem[{Plait {et~al.}(1995)Plait, Lundqvist, Chevalier, \&
  Kirshner}]{Plait:1995p25}
Plait, P.~C., Lundqvist, P., Chevalier, R.~A., \& Kirshner, R.~P. 1995, \apj, 439, 730

\bibitem[{Podsiadlowski {et~al.}(2007)Podsiadlowski, Morris, \&
  Ivanova}]{Podsiadlowski:2007p9791}
Podsiadlowski, P., Morris, T.~S., \& Ivanova, N. 2007, SUPERNOVA 1987A: 20
  YEARS AFTER: Supernovae and Gamma-Ray Bursters. AIP Conference Proceedings,
  937, 125, (c) 2007: American Institute of Physics

\bibitem[{{Pun}(2007)}]{pun:2007}
{Pun}, C.~S.~J. 2007, in American Institute of Physics Conference Series, Vol.
  937, Supernova 1987A: 20 Years After: Supernovae and Gamma-Ray Bursters, ed.
  S.~{Immler}, K.~{Weiler}, \& R.~{McCray}, 171--175

\bibitem[{Reynolds {et~al.}(1995)Reynolds, Jauncey, Staveley-Smith, Tzioumis,
  de~Vegt, Zacharias, Perryman, van Leeuwen, King, McCulloch, Russell,
  Johnston, Hindsley, Malin, Argue, Manchester, Kesteven, White, \&
  Jones}]{Reynolds:1995p3810}
Reynolds, J.~E., {et~al.} 1995, Astronomy and Astrophysics, 304, 116

\bibitem[{Reynolds {et~al.}(2008)Reynolds, Borkowski, Green, Hwang, Harrus, \&
  Petre}]{Reynolds:2008p14576}
Reynolds, S.~P., Borkowski, K.~J., Green, D.~A., Hwang, U., Harrus, I., \&
  Petre, R. 2008, \apj, 680, L41, (c) 2008: The American
  Astronomical Society

\bibitem[{{Richardson}(1972)}]{richardson:1972}
{Richardson}, W.~H. 1972, Journal of the Optical Society of America
  (1917-1983), 62, 55

\bibitem[{Sault {et~al.}(1995)Sault, Teuben, \& Wright}]{Sault:1995p14420}
Sault, R.~J., Teuben, P.~J., \& Wright, M. C.~H. 1995, Astronomical Data
  Analysis Software and Systems IV, 77, 433

\bibitem[{Schure {et~al.}(2008)Schure, Vink, Achterberg, \&
  Keppens}]{Schure:2008p12137}
Schure, K., Vink, J., Achterberg, B., \& Keppens, R. 2008, 37th COSPAR
  Scientific Assembly. Held 13-20 July 2008, 37, 2791

\bibitem[{Staveley-Smith {et~al.}(2007)Staveley-Smith, Gaensler, Manchester,
  Ball, Kesteven, \& Tzioumis}]{StaveleySmith:2007p2}
Staveley-Smith, L., Gaensler, B.~M., Manchester, R.~N., Ball, L., Kesteven,
  M.~J., \& Tzioumis, A.~K. 2007, SUPERNOVA 1987A: 20 YEARS AFTER: Supernovae
  and Gamma-Ray Bursters. AIP Conference Proceedings, 937, 96

\bibitem[{Staveley-Smith {et~al.}(1992)Staveley-Smith, Manchester, Kesteven,
  Reynolds, Tzioumis, Killeen, Jauncey, Campbell-Wilson, Crawford, \&
  Turtle}]{StaveleySmith:1992p680}
Staveley-Smith, L., {et~al.} 1992, Nature (ISSN 0028-0836), 355, 147

\bibitem[{Sugerman {et~al.}(2005)Sugerman, Crotts, Kunkel, Heathcote, \&
  Lawrence}]{Sugerman:2005p11362}
Sugerman, B. E.~K., Crotts, A. P.~S., Kunkel, W.~E., Heathcote, S.~R., \&
  Lawrence, S.~S. 2005, \apjs, 159, 60,
  (c) 2005: The American Astronomical Society

\bibitem[{Tanaka \& Washimi(2002)}]{Tanaka:2002p19}
Tanaka, T., \& Washimi, H. 2002, Science

\bibitem[{Taylor {et~al.}(1999)Taylor, Carrilli, \& Perley}]{ekers:1999}
Taylor, G., Carrilli, C., \& Perley, R., eds. 1999, Ekers R.~D, in ASP Conf.
  Ser. 180, Synthesis Imaging in Radio Astronomy (San Francisco: San Francisco:
  ASP)

\bibitem[{Tingay {et~al.}(2009) Tingay, Philips, Amy, Tzioumis, Kettenis, Boven, Szomoru, Paragi, van Langevelde, Verkouter, Philips, Cowie, Tam, Huisman }]{tingay:2009} Tingay, S.J., Philips, C.J., Amy S.W., Tzioumis, A.K., Kettenis, M., Boven, E.P., Szomoru, A., Paragi, Z., van Langevelde, H.J., Verkouter, H., Philips, I., Cowie, A., Tam, T., Huisman, W., proceedings of the The 8th International e-VLBI Workshop, Proceedings of Science 

\bibitem[{Trimble(1973)}]{Trimble:1973p13894}
Trimble, V. 1973, \pasa, 85,
  579, a{\&}AA ID. AAA010.134.010

\bibitem[{Tsunemi {et~al.}(2001)Tsunemi, Mori, Miyata, Baluta, Burrows,
  Garmire, \& Chartas}]{Tsunemi:2001p8231}
Tsunemi, H., Mori, K., Miyata, E., Baluta, C., Burrows, D.~N., Garmire, G.~P.,
  \& Chartas, G. 2001, \apj, 554, 496, (c) 2001: The
  American Astronomical Society

\bibitem[{Turtle {et~al.}(1987)Turtle, Campbell-Wilson, Bunton, Jauncey, \&
  Kesteven}]{Turtle:1987p2190}
Turtle, A.~J., Campbell-Wilson, D., Bunton, J.~D., Jauncey, D.~L., \& Kesteven,
  M.~J. 1987, Nature (ISSN 0028-0836), 327, 38

\bibitem[{Zanardo {et~al.}(2009)Zanardo, Staveley-Smith, Manchester, Ng,
  Gaensler, Ball, \& Kesteven}]{zanardo:2009}
Zanardo, G., Staveley-Smith, L., Manchester, R.~N., Ng, C.-Y., Gaensler, B.,
  Ball, L., \& Kesteven, M. 2009, in-preparation

\end{thebibliography}
\end{document}